\documentclass[fleqn,usenatbib,useAMS]{mnras}

\usepackage{graphicx}	
\usepackage{amsmath}	
\usepackage{amssymb}	
\usepackage{multicol}        
\usepackage{bm}		
\usepackage{pdflscape}	

\usepackage[T1]{fontenc}
\usepackage{ae,aecompl}

\usepackage{newtxtext,newtxmath}
\usepackage{color}

\title[Clustering in EN1 field]{{The study of the angular and spatial distribution of radio selected AGNs and star-forming galaxies in the ELAIS N1 field}}

\author[Arnab Chakraborty]{Arnab Chakraborty,$^{1}$\thanks{\href{mailto:phd1601121009@iiti.ac.in}{phd1601121009@iiti.ac.in}}
Prasun Dutta, $^{2}$
Abhirup Datta,$^{1}$
Nirupam Roy, $^{3}$
\\
$^{1}$Discipline of Astronomy, Astrophysics and Space Engineering, Indian Institute of Technology Indore, Indore 453552, India\\
$^{2}$Department of Physics, IIT (BHU) Varanasi, 221005 India\\
$^{3}$Department of Physics , Indian Institute of Science,Bangalore 560012,India\\}
\date{Last updated 2015 May 22; in original form 2013 September 5}

\pubyear{2015}

\begin{document}
\label{firstpage}
\pagerange{\pageref{firstpage}--\pageref{lastpage}}
\maketitle

\begin{abstract}
The cosmic evolution of bias of different source populations with underlying dark matter density field in post reionization era can shed light on large scale structures.  Studying  the angular and spatial distribution of different compact sources using deep radio catalogue at low-frequency is essential to understand the matter distribution of the present Universe. Here, we investigate the relationship of luminous matter with their host dark matter haloes by measuring  the angular and spatial clustering of sources (two-point statistics), using deep radio observation of ELAIS N1 (EN1) field with upgraded Giant Metrewave Radio Telescope (uGMRT) at 300-500 MHz. We also analyze the 612 MHz GMRT archival data of the same field to understand the cosmic evolution of clustering of different source populations. We classify the sources as star-forming galaxies (SFGs) and active galactic nuclei (AGN) based on their radio luminosity.   We find that the  spatial clustering length and bias to the dark matter density field of  SFGs are smaller than AGNs at both frequencies. This proves that AGNs are mainly hosted by massive haloes and hence strongly clustered. However, a small decrease in the bias for both kind of sources at higher frequency indicates  that we are most likely tracing the faint objects residing in less massive haloes at higher frequencies. Our results are in excellent agreement with previous findings at radio and multi-frequency surveys. However, comparison  with SKADS simulation suggests that the halo mass for different populations  used in the simulation is systematically lower.  This work  quantifies the spatial distribution of extragalactic compact objects in EN1 field and bridges the gap between shallow  and deep surveys.

\end{abstract}





\section{Introduction}
Observation of the oldest light of the Universe, Cosmic Microwave Background (CMB) radiation, have shown that the early Universe (at $\textit{z} \sim 1100$) was remarkably smooth  with tiny anisotropies of the order of $10^{-4} $\citep{Planck2016}. The Universe was almost neutral at that epoch. The linear fluctuations in the matter density  grew under gravitational clustering and eventually generated luminous objects. The high energy photons emanating from the first stars and galaxies started heating and ionizing the intergalactic medium (IGM) and forced the Universe to go through a global phase transition.  Constraint from Gunn Peterson trough in quasar absorption spectra indicate that the gas in the intergalactic medium (IGM) became completely ionized by redshift $\textit{z} \sim 6$ \citep{Fan2002,Fan2006}. The transition of the universe from the neutral to ionized phase is called the epoch of reionization (EoR) \citep{Field1958,Madau1997,Furlanetto2006,Pritchard2012,Loeb2013,Mesinger2016,Barkana2016,Dayal2018}). In the post reionization universe, several astrophysical parameters other than the gravitational clustering influenced the evolution of the baryonic matter density. Estimation of the clustering pattern of different radio sources and comparing it to the dark matter power spectrum holds the key to understand the astrophysical aspects of the post EoR evolution of the matter density distribution.

 The extragalactic sky at low-frequency in $\mu$Jy to mJy flux limit is mainly dominated by Fanaroff–Riley type I (FR I), radio quiet quasars and star-forming galaxies. One of the possible way to  understand these sources along with their host dark matter haloes  is through their clustering. The clustering of galaxies can be quantified through counting excess number of  galaxies at a certain scale to that of a random distribution of galaxies. This  is known as the  two-point correlation function \citep{Lindsay2014,Hale2018}. Traditionally,  the two-point angular correlation function using optical galaxy surveys are studied and  the redshift information is used to estimate the spatial clustering and the bias with the underlying matter power spectrum. However, covering a large fraction of sky is extremely difficult in the optical surveys. Furthermore, dust obscuration limits the  detection of high redshifts sources. Emissions in radio wavelengths from these galaxies are not  attenuated by dust and hence provide a direct probe of high $\textit{z}$ galaxies \citep{Jarvis2016}. The relatively larger field of view of the radio telescopes also increase the survey speed. So, deep radio surveys allow to estimate the cross-correlation of the luminous matter distribution with CMB and shed light on how the baryons  traces the underlying dark matter distribution \citep{Planck2014,Allison2015}.  Multifrequency observations can be used to understand the relationship between AGNs and star formation processes with their host dark matter haloes \citep{Hale2018}. Previous findings suggest that AGNs are mainly hosted by massive haloes and strongly clustered than SFGs \citep{Gilli2009,Donoso2014,Magliocchetti2017,Hale2018}. An efficient method to understand the bias of these populations to the underlying dark matter density field is the ratio of spatial correlation lengths of luminous matter and dark matter density field \citep{Peebles1980}.  This study of large scale distribution of dark matter through clustering pattern of luminous objects is important to understand the structure formation \citep{Hale2018}.  Measurement of bias for different populations of sources in radio surveys allow to probe non-Gaussianity in the initial density fluctuation \citep{Seljak2009,Raccanelli2015}.

The 21 cm signal from the EoR and post EoR universe traces the evolution of the matter density over all redshifts. The major hindrance on our quest to detect the redshifted 21 cm signal  are bright dominating foregrounds.  Foreground contamination due to diffuse Galactic synchrotron emission from our Galaxy (DGSE) \citep{Shaver1999}, free-free emission from Galactic and extragalactic sources \citep{Cooray2004}, faint radio-loud quasars \citep{Di Matteo2002}, synchrotron emission from low-redshift Galaxy clusters \citep{Di Matteo2004}, extragalactic point sources \citep{Di Matteo2004}, etc,  are several order of magnitudes higher than the cosmological signal of interest \citep{Zaldarriag2004,Bharadwaj2005,Jeli&cacute;2008,Bernardi2009,Jeli&cacute;2010,Zaroubi2012}.  \citet{Di Matteo2004} have shown that spatial clustering of extragalactic sources with flux densities $\gtrsim $0.1 mJy  dominates the angular fluctuation at scales, $\theta \gtrsim$ 1 arcmin and modelling and removing these sources will allow us to detect angular fluctuation in 21 cm emission. However, to understand the systematic effects of the foregrounds on calibration and the effect of bright  point sources far away from the field of view of observations, we require an accurate statistical model of the compact source distribution. 
 The extra-galactic source counts are often modelled as single power law  or a smooth polynomial \citep{Intema2017,Franzen2019}  and the spatial distribution of sources is assumed to be Poissonian  \citep{Ali2008,Trott2016} or a simple power-law clustering pattern has been implemented.  The assumption of simple Poissonian  distribution of sources in EoR foreground modelling is too simplistic and will effect the signal extraction gravely \citep{Ali2008, Jeli&cacute;2010, Ghosh2012,Trott2016}. This demands a better observational estimate of the clustering pattern of the point sources at the lower radio frequencies.

With this motivation, here we have studied in detail the clustering pattern of different sources (AGNs and SFGs)  in EN1 field using deep radio survey  with uGMRT,  at 300 - 500 MHz. Along with that, to understand the redshift evolution of bias of different sources we also study the  clustering of different populations using 612 MHz GMRT archival data of the same field. In our previous works, we have studied the spatial and spectral properties of diffuse galactic synchrotron emission, the angular power spectrum of Galactic and extragalactic foregrounds, the  differential source counts as a function of flux densities, etc in the same field \citep{Chakraborty2019A, Chakraborty2019B}. 

The paper is structured as follows.  The details of radio and optical data used in this work is presented in Sec. \ref{Data}. The pre-processing of the source catalogues to estimate the angular and spatial clustering of sources  and the classification of sources based on radio luminoisty is discussed in Sec. \ref{preprocessing}. The estimated angular correlation of all sources and a detailed comparison with previous findings are presented in Sec. \ref{sec_angular}. The results of spatial clustering and bias of the sources to the dark matter density field is discussed in Sec. \ref{sec_spatial}. The clustering pattern of AGNs and SFGs along with the detail comparison with previous observations and simulations are discussed in Sec. \ref{sec_AGN_SFG}.  Finally, we draw conclusion in Sec. \ref{sec_conclusion}.  

Throughout this work, we use the best fitted cosmological parameters of the Planck 2018 analysis \citep{Planck2018}, which are: $\Omega_{\mathrm{M}}$ = 0.31, $\Omega_{\Lambda}$ = 0.68, $\sigma_{8}$ = 0.811, $H_{0}$ = 67.36 Km $\mathrm{s}^{-1}$ $\mathrm{MPc}^{-1}$.

\section{Sources of Data at different frequency bands}
\label{Data}

We use 400 MHz uGMRT observation of EN1 field along with archival GMRT 612 MHz data for cataloguing radio sources at these frequency bands. 
In this section, we briefly discuss the radio and optical data used in this work to estimate the angular and spatial clustering properties of sources in the EN1 field.  The details of radio surveys are mentioned in Table \ref{survey_paramateres}.

 \begin{table}
\caption{Details of different parameters of radio data used in this work}
\label{survey_paramateres}
\scalebox{1.0}{
\begin{tabular}{|c|c|c|c|c|c|c|c|}
\hline
\hline
Target & $\nu$ &   $S_{\mathrm{limit,\nu}}$ &  Survey area &  Total sources \\

&  (MHz) &   ($\mu$Jy beam$^{-1}$) & (deg$^{2}$)& (N)  \\ 
\hline
EN1 & 400 &  100.0 & 1.8 & 2528  \\
\hline
EN1 & 612 &  50.0 & 1.13 & 2342 \\

\hline
\end{tabular}}
\flushleft{${\dagger}$ $S_{\mathrm{limit,\nu}}$ is the flux density limit of the corresponding catalogue at observed frequency ($\nu$). \\
 }
\end{table}

\subsection{uGMRT observation at 400 MHz}
 The EN1 field ($\alpha_{2000}=16^{h}10^{m}1^{s} ,\delta_{2000}=54^{\circ}30'36\arcsec$ ) has been observed  with uGMRT in GTAC cycle 32 during May 2017 at 300 - 500 MHz for a total on source  time of 25  hours.  The raw data  has a time resolution of $2$ sec and frequency resolution of $24$ KHz.  This high resolution data was flagged for RFI  with AOFLAGGER \citep{Offringa2012} and then averaged  to 16 sec time  and 97 kHz frequency resolution for further processing. The calibration, imaging and self-calibration  was done in  {\tiny CASA} \footnote{See: \url{https://casa.nrao.edu}; \citep{McMullin2007} } \citep{Chakraborty2019B}. The final image, covering $\sim$ 1.8 deg$^2$ area, has a  off-source noise of $\sim 15\mu \mathrm{Jy}$ $\mathrm{beam^{-1}}$ in the central part ( Fig. \ref{image_400}: left panel). A  source catalogue of  2528 sources with flux densities above 100 $\mu$Jy (>6$\sigma$)  was compiled using P{\tiny Y}BDSF \citep{Mohan2015}. In this work we use these catalogued compact radio sources in the EN1 field  to estimate the angular and spatial clustering (the catalogue is publicly available \footnote{\url{http://vizier.u-strasbg.fr/viz-bin/VizieR?-source=J/MNRAS/490/24}}) . The detail analysis of the uGMRT $400$ MHz observation, calibration, source extraction and catalogue formation is presented in \citet{Chakraborty2019B}.

 \subsection{GMRT archival data at 612 MHz}
\begin{figure*}
\centering
\begin{tabular}{c|c}
\includegraphics[width=\columnwidth,height=3in]{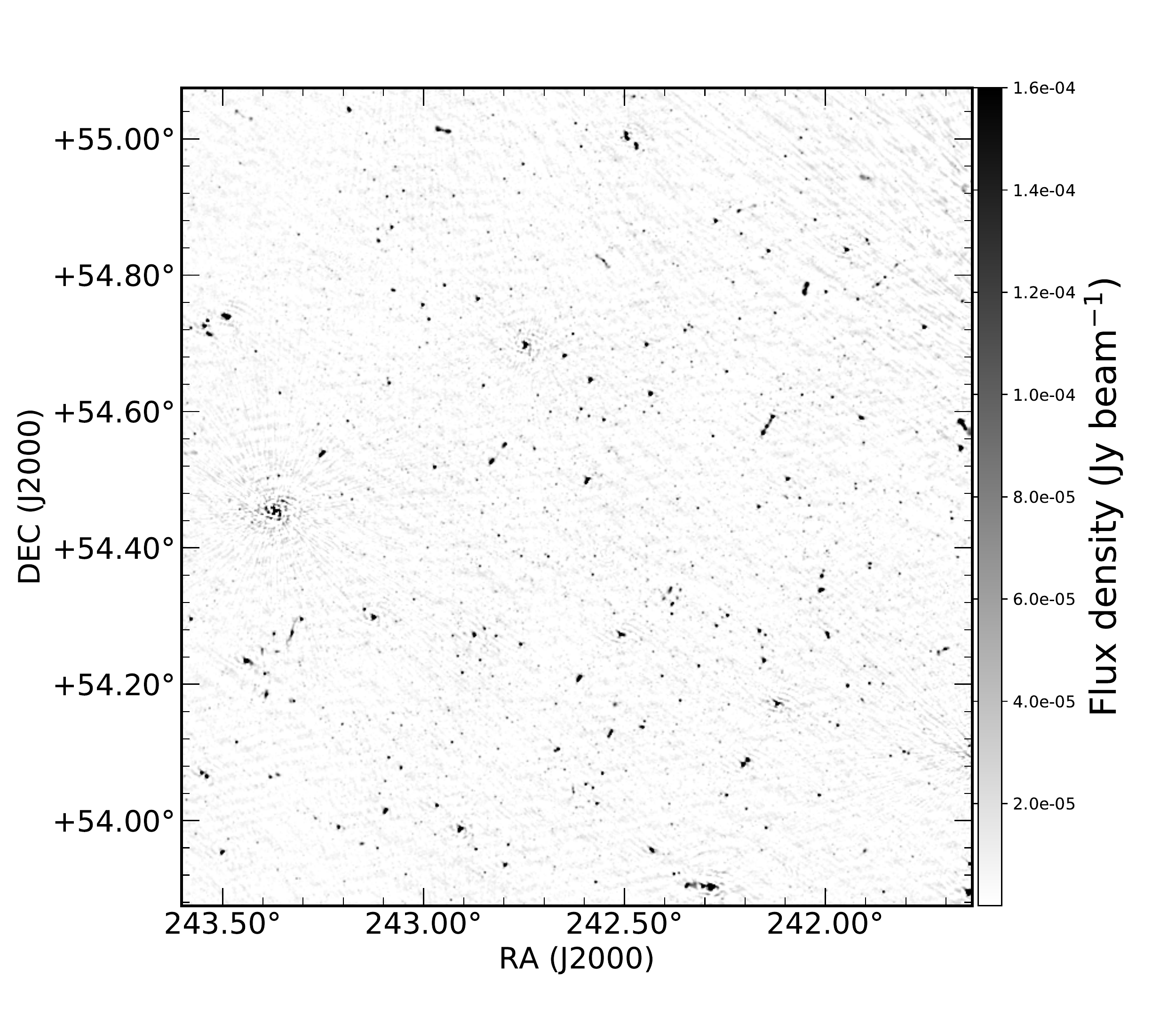} &
\includegraphics[width=\columnwidth,height=3in]{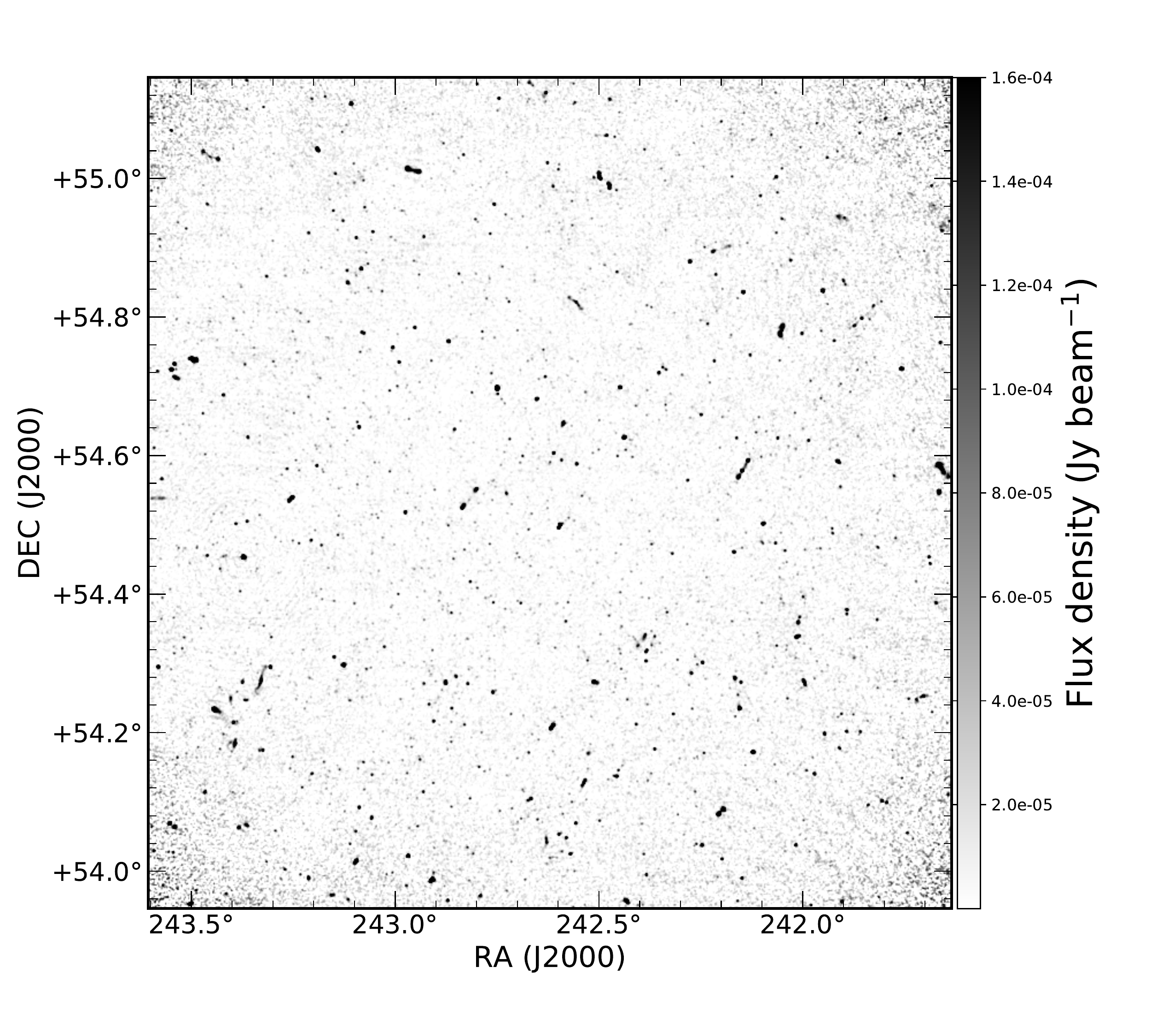}
\end{tabular}
\caption{The central 1.2 $\mathrm{deg}^{2}$ of  total intensity map of the ELIAS N1 at 400MHz (left) and 612 MHz (right) are shown. The off-source rms at the center is $\sim$ 15 $\mu \mathrm{Jy}$ $\mathrm{beam^{-1}}$ for 400 MHz map and   $\sim$ 8 $\mu \mathrm{Jy}$ $\mathrm{beam^{-1}}$ for 612 MHz map.}
\label{image_400}
\end{figure*}

 We additionally use the GMRT archival data of EN1 field, observed between 2011 to 2013 at 612 MHz (project codes: 20\_044, 21\_083, 22\_056) in this study.  We re-analyze this GMRT archival data and briefly mention the procedure here. Observations were carried out  using seven different pointings arranged in a hexagonal pattern centered at $\alpha_{2000}=16^{h}10^{m}30^{s}. ,\delta_{2000}=54^{\circ}35'00\arcsec$ \citep{Taylor2016}. We  perform necessary flagging and calibration of each pointing using the AIPS based automated pipeline Source peeling and Atmospheric Modeling (SPAM)  \citep{Intema2009,Intema2017}.  We  create a  mosaic of the seven  pointings to create a combined image of the field which covers $\sim$ 1.13 deg$^2$ area. We achieve a  off-source rms of $\sim 8\mu \mathrm{Jy}$ $\mathrm{beam^{-1}}$ at the central part of the image (Fig. \ref{image_400}: right panel). We use P{\tiny Y}BDSF on the final image to compile a 612 MHz  source catalogue of the EN1 field having 2342 sources  with flux densities above 50 $\mu$Jy ($>6\sigma$).  
 
\subsection{BOSS spectroscopic survey}

The  Baryon Oscillation Spectroscopic Survey (BOSS) is the dark time survey of the third phase of the Sloan Digital Sky Survey or SDSS-III (SDSS; \citealt{York2000,Dawson2013,Alam2015}). This survey primarily aims to determine Baryon Acoustic Oscillation feature  (BAO) in large scale structure \citep{Cole2005,Eisenstein2005}. It has measured redshift of $1.5$ million galaxies and have obtained distance-redshift relation $d_{A}(z)$ and the Hubble parameter $H(z)$ with highest precision till date. 
Four plates were granted  to observe spectra of the radio sources identified in the EN1 field by LOFAR, GMRT, JVLA and FIRST. After data reduction, spectra of sources observed in each plate along with their position and redshift is made available as a part of SDSS DR12. We use this  spectroscopic redshift catalogue released by SDSS team to identify the redshift of the radio sources in our catalogues discussed above \footnote{\url{https://www.sdss.org/dr12/algorithms/ancillary/boss/sdsslofar/}} \citep{Alam2015}.

\subsection{SWIRE photometric survey}
 The revised Spitzer Wide-Area Infrared Extragalactic survey (SWIRE)  catalogue   incorporates   Two  Micron  All  Sky  Survey  (2MASS)  and  UKIRT  Infrared  Deep  SkySurvey (UKIDSS) near-infrared data, which essentially reduces the fraction of catastrophic outliers  and gives a more reliable photometric catalogue \citep{Rowan-Robinson2008,Rowan-Robinson2013}. The SWIRE survey covers a sky of 49 $\mathrm{deg^{2}}$ at 3.6, 4.5, 5.8, 8.0, 24.0, 70.0 and 160.0 $\mu m$ using Spitzer. This survey covers $\sim$ $8.72$ $\mathrm{deg^{2}}$ of EN1 field in five bands ($U\arcmin,g\arcmin,r\arcmin,i\arcmin,Z\arcmin$). Together with this, the additional information of JHK magnitude from  2MASS and UKIDSS results in the most reliable photometric redshift catalogue publicly available for this field. 
 
 We use the spectroscopic redshifts in the SDSS catalogue and the photometric redshifts in the  SWIRE catalogue complimentarily to get redshifts of the sources identified in the radio frequencies.

\subsection{Semi-empirical extra galactic radio source catalogue}

\citet{Wilman2008} performed semi-empirical simulation to generate extragalactic radio continuum sky where they model the statistical distribution of different sources to be seen in radio continuum with the next generation radio telescope. Their conical simulation volume covers a sky area of $20 \times 20$ deg$^{2}$ to a redshift of $z=20$ with sources to a flux density limit of $10$ nJy.  They use five distinct source types, the radio quiet quasars (RQQ), the radio loud AGNs with FRI and FRII morphology, the star forming galaxies with quiescent  and starbursts (SB). They use a numerical Press-Schecter \citep{Press1974} formalism to identify high density halos in the dark matter distribution. A source type is then identified with the given halo mass and the luminosity of the source  drawn from an observed luminosity function of the given source type. In this work we use the corresponding source catalogue, statistical inference of the variation of angular and spatial clustering of different source types over the redshift. We shall use $S^{3}$ simulation or $S^{3}$ (SKA Simulated Skies) to denote results of this simulation henceforth.

\section{Preprocessing of the source catalogue}
\label{preprocessing}
To generate radio catalogue at 400 MHz and 612 MHz with redshift information and source characteristics we proceed with the steps given in this section. These catalogues are used to estimate the angular and spatial clustering of the radio sources.

\subsection{Merging multi-component sources}

In a high resolution radio map, such as here ($\sim$ 5$\arcsec$),  many sources  resolve into multiple components. These are mostly radio galaxies with a  central nucleus; accompanied by hotspots along or at the end of, one or two jets \citep{Prandoni2018}. As a result, a single radio source may split into multiple  sources in the catalogue \citep{Magliocchetti1998}. It is important to identify these separate components  as a single source to unbiasedly  quantify the angular clustering.  \citet{Oort1987} finds a strong correlation between the angular size $\theta$ of the radio sources with their flux densities $S$. This relation, $\theta \propto \sqrt{S}$, usually known as the  $\theta$ -$S$ relation, is used to identify multiple resolved components of radio sources in various earlier surveys \citep{Magliocchetti1998}. 

\begin{figure*}
    \centering
    \begin{tabular}{c|c}
    \includegraphics[width=\columnwidth]{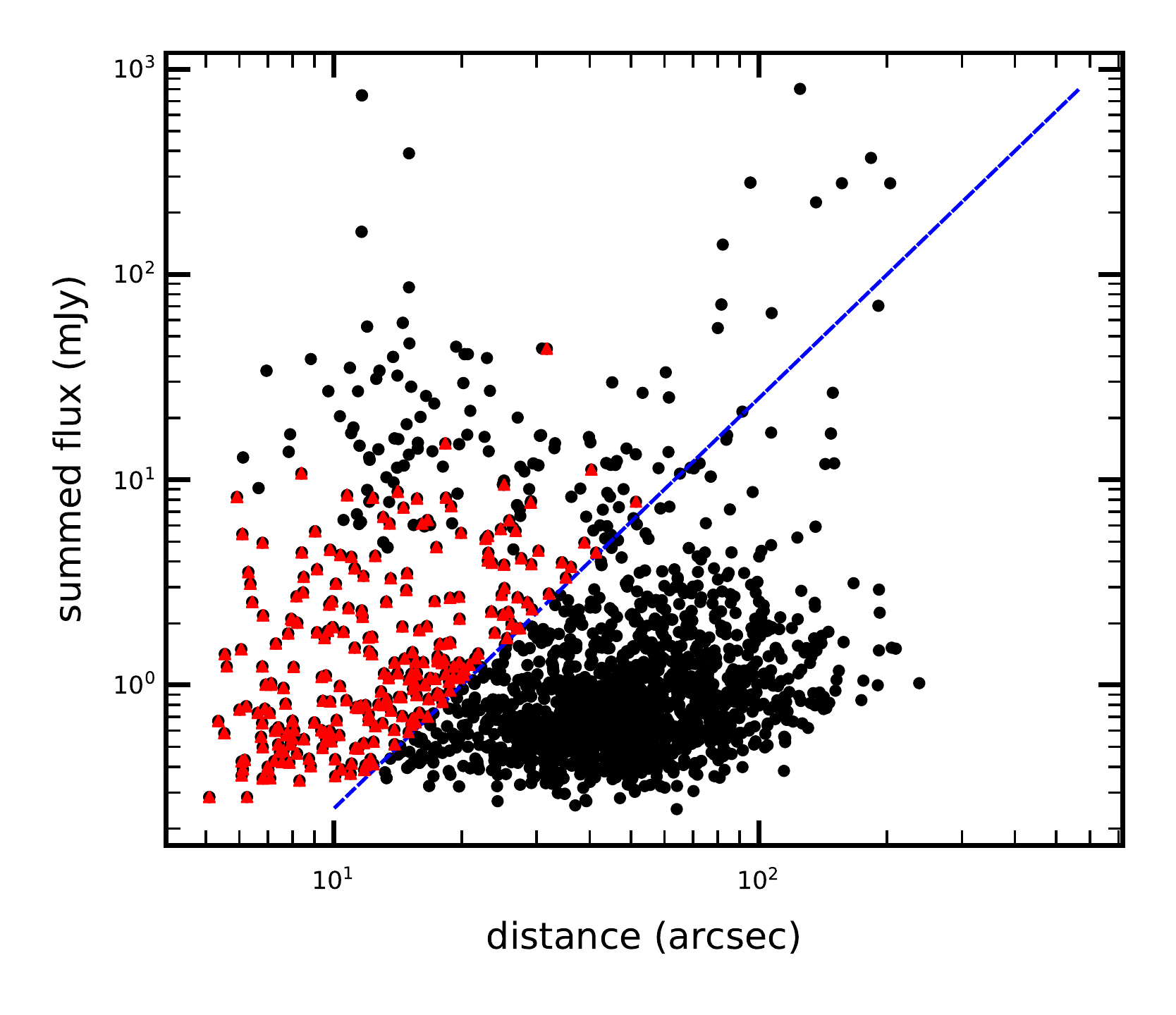}&
    \includegraphics[width=\columnwidth]{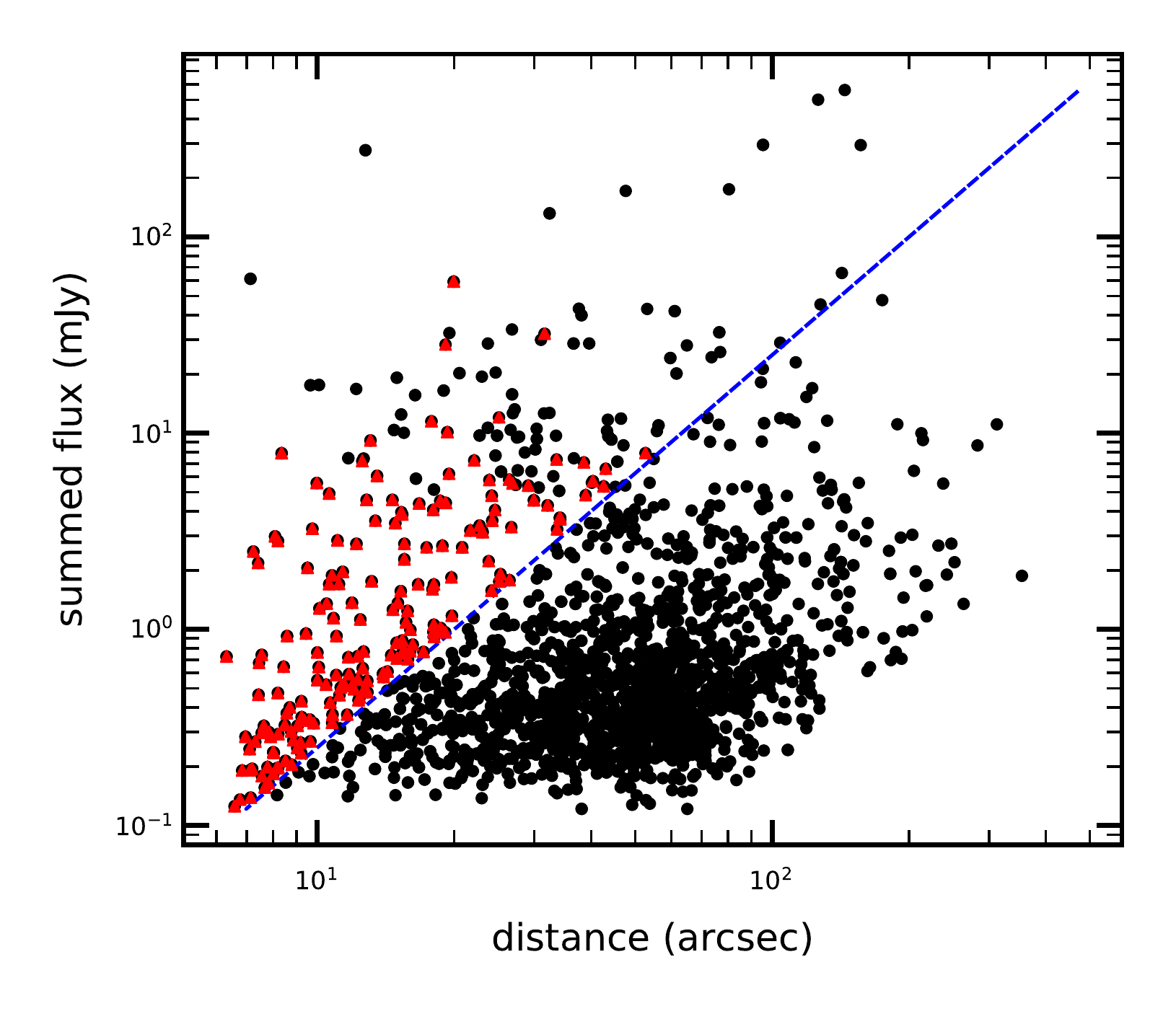}
    \end{tabular}
    \caption{Sum of flux densities of nearest neighbours pairs as a function of their angular separation is plotted in black circles. The pair of sources  lying above the blue dashed line and with flux density difference  less than by a factor of 4 are shown in red. Left panel and right panel are for 400 MHz and 612 MHz catalogue, respectively.}
    \label{double_400}
\end{figure*}

 We use the following two criteria to associate multiple nearby objects in our initial catalogue  as a single source in the final catalogue. We first estimate the total flux $S_{\mathrm{total}}$ and angular separations $\theta$ of each pairs of sources and identify the pairs to a maximum angular separation $\theta_{\mathrm{max}}$. The choice of maximum separation is bard on the $\theta$ -$S$ relation discussed above and is given as $\theta_{\mathrm{max}} = 20(S_{\mathrm{total}})^{0.5}$, where $\theta_{\mathrm{max}}$ is given in arcsec and $S_{\mathrm{total}}$ in mJy \citep{Huynh2005,Prandoni2018}. We consider the above identified pairs to come from a single source if their flux densities  differ by a factor of less than $4$ \citep{Huynh2005}. Fig. \ref{double_400} shows the summed flux of each nearest pair as a function of distance between them in black filled circles. The pairs above the blue dashed line have angular separation lesser than $\theta_{\mathrm{max}}$. Sources satisfying both the criteria discussed above 
are shown as red triangles and are considered as single sources in the final catalogue. Using this method  we have  2253 and 2153 sources in our final radio catalogue  at 400 MHz and 612 MHz respectively.

\subsection{Adding Redshift information of sources} 
\label{redshift_info}
We use the BOSS spectroscopic survey and the SWIRE photometric survey for optical identification of sources  in our radio catalogue. The positional accuracy of the uGMRT observation  is less than 1$\arcsec$ \citep{Chakraborty2019B}. Hence, we use a simple nearest-neighbour match between the sources in optical and radio catalogue.  The search is performed within a radius of $r_{\mathrm{s}}$ from the sources. \citet{Lindsay2014} show that the rate of contamination $P_{\mathrm{c}}$ of the  remaining sources due to closeness to optical sources can be expressed as:
\begin{equation}
P_{\mathrm{c}} =  \pi r_{\mathrm{s}}^{2} \sigma_{\mathrm{opt}},  
\end{equation}
where $r_{\mathrm{s}}$ is the search radius cut-off and $\sigma_{\mathrm{opt}}$ is the surface density of optical catalogue. We have chosen $r_{\mathrm{s}}$ as $3\arcsec$.  For $\sigma_{\mathrm{opt}} = 1.4 \times 10^{4} \mathrm{deg}^{-2}$,  a $3\arcsec$ search radius results in a contamination rate of 3.1\%. This search radius allows us to include a large number of radio samples as well as ensures valid optical identification. We use this search radius for cross-matching. 
\begin{figure*}
 
   \centering
\begin{tabular}{c|c}
      
    \includegraphics[width=\columnwidth,height=3in]{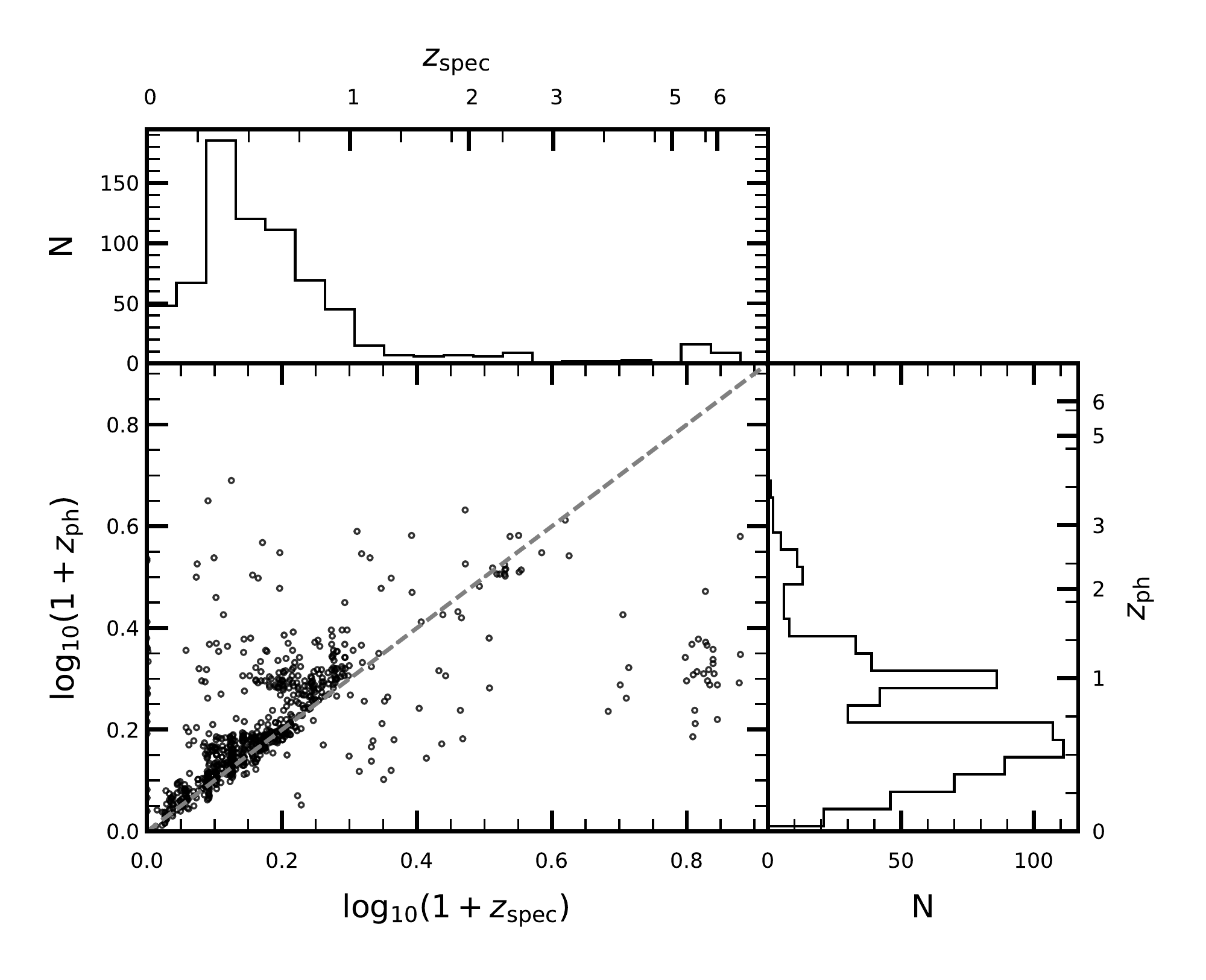} &
    \includegraphics[width=\columnwidth,height=3in]{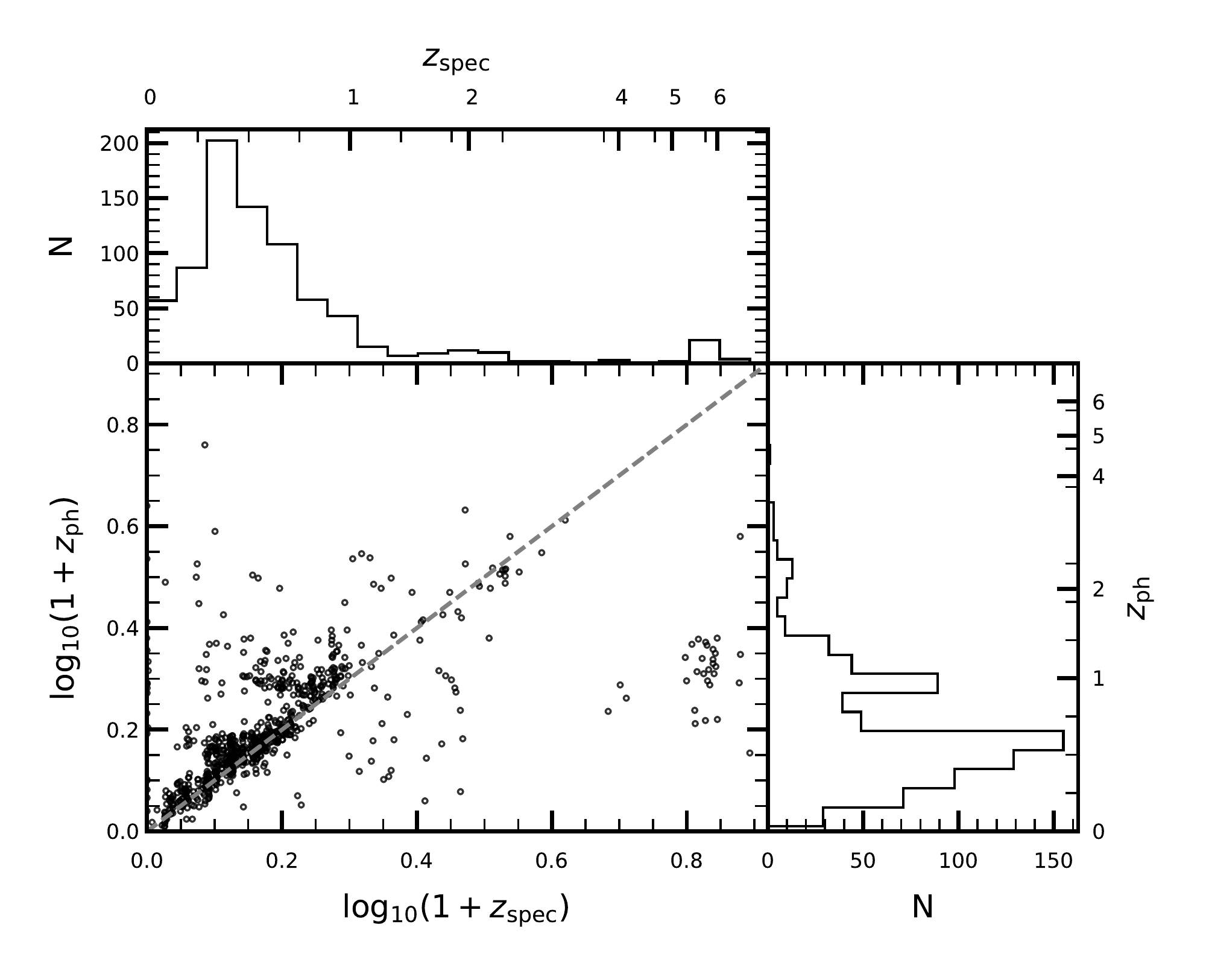}
    \end{tabular}
    \caption{The comparison between photometric and spectroscopic redshift of sources with both for 400 MHz (left panel) and 612 MHz (right panel) catalogue. The histogram of redshifts are also plotted here. The redshift distribution of sources for both the catalogues are nearly identical.}
    \label{redshift_hist}
\end{figure*}

We find 761 (30\%) and 822 (35\%) sources with spectroscopic redshifts  and 1470 (58\%) and 1391 (59\%) sources with photometric redshifts  for the 400 MHz and 612 MHz catalogues respectively. Fig. \ref{redshift_hist} shows the correlation plots to compare the photometric and spectroscopic redshifs for both the catalogues respectively. Note that, spectroscopic and photometric redshifts are consistent  for low redshift sources for both the catalogues. We also show the histogram of spectroscopic and photometric redshifts for both the catalogues in Fig \ref{redshift_hist}. We associate redshift information to a total of  1503 (59\%) sources for 400 MHz catalogue and 1429 (61\%) sources for 612 MHz catalogue. We use photometric redshift only for those sources, which do not have spectroscopic redshift information available.

\begin{figure*}
 
   \centering
\begin{tabular}{c|c}
      
    \includegraphics[width=\columnwidth,height=3in]{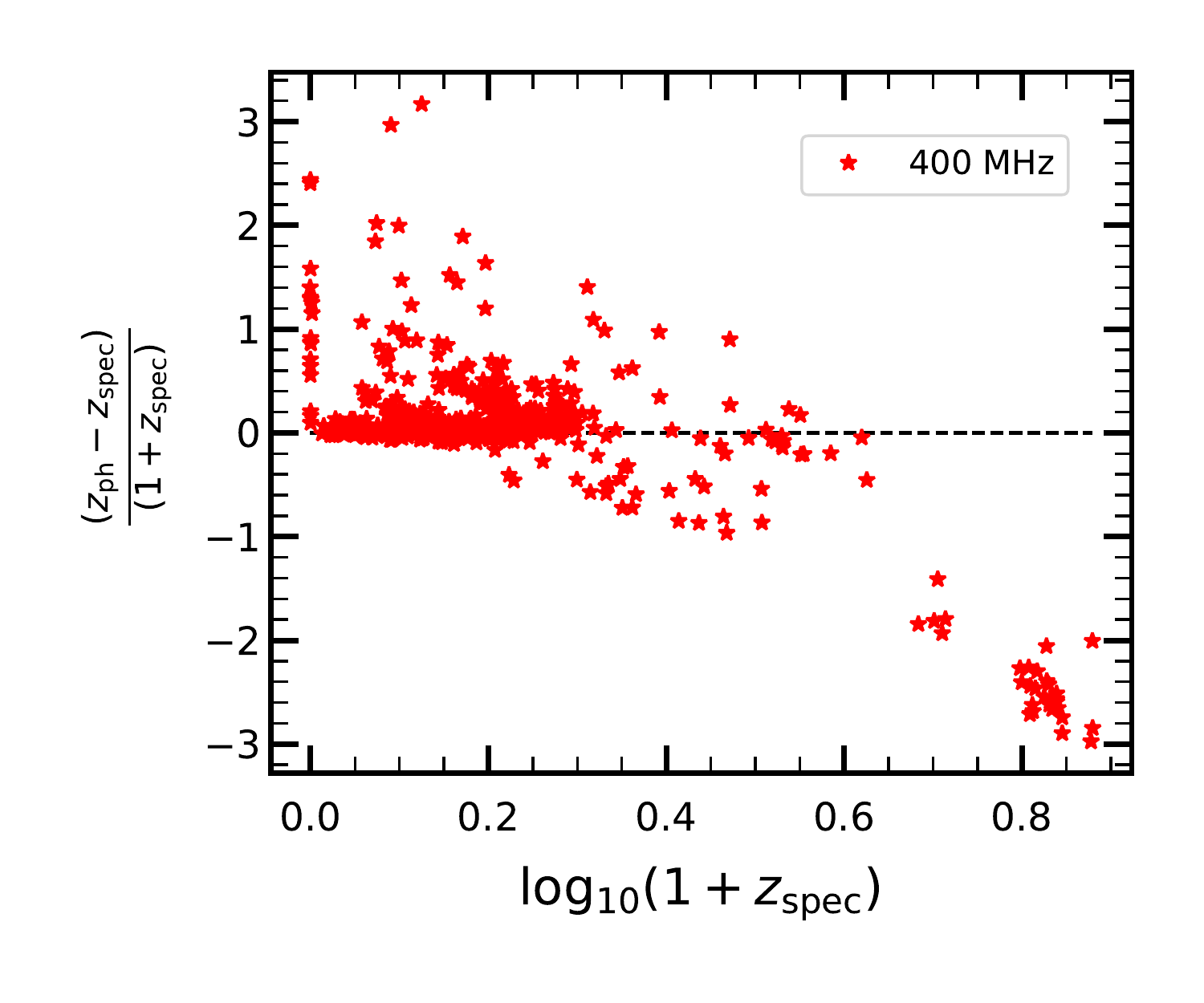} &
    \includegraphics[width=\columnwidth,height=3in]{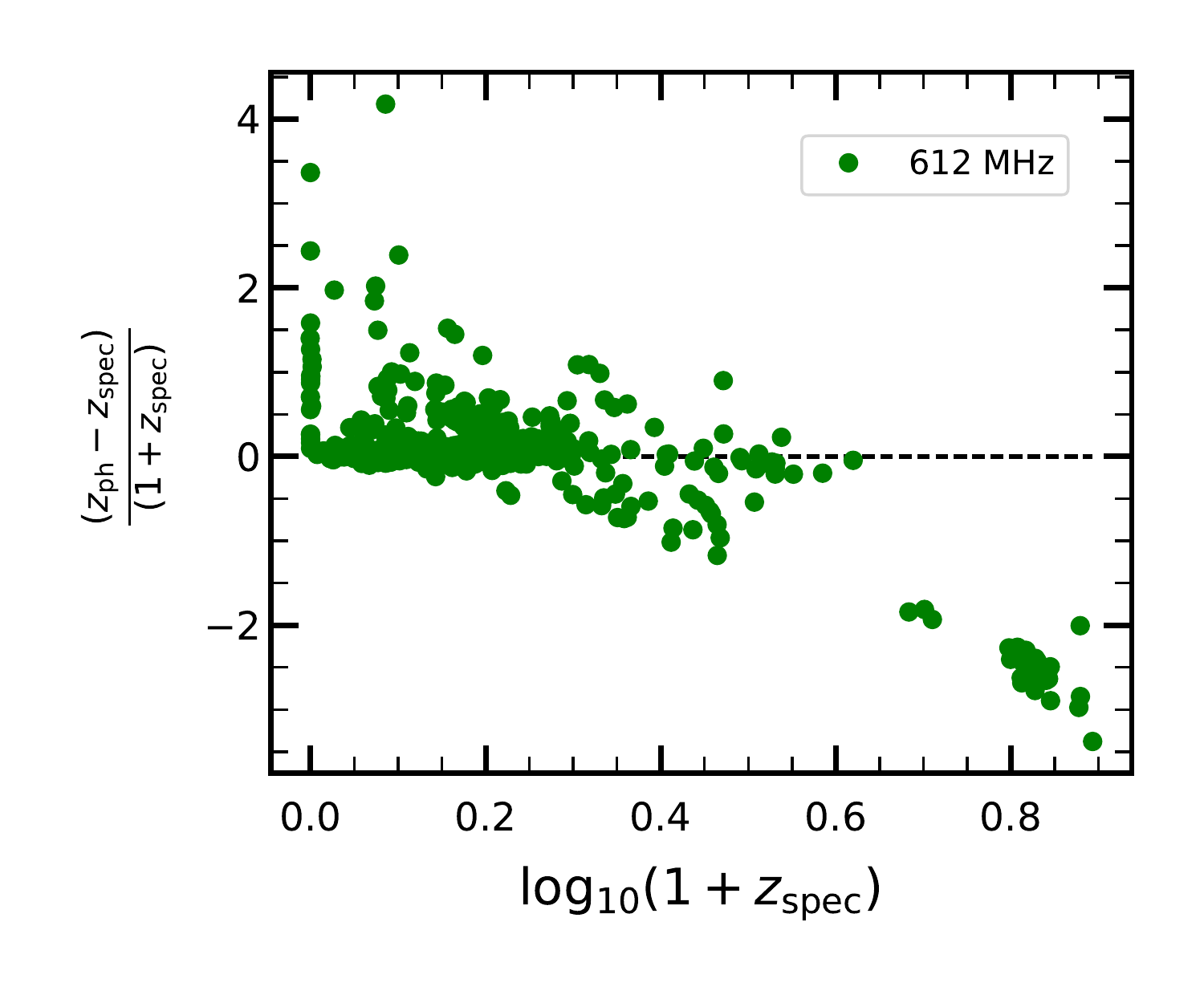}
    \end{tabular}
    \caption{The figure shows the distribution $\frac{(z_{\mathrm{ph}} - z_{\mathrm{spec}})}{(1+z_{\mathrm{spec}})}$ as a function of $z_{\mathrm{spec}}$. The black dashed line represent zero deviation of photometric redshift with respect to spectroscopic redshift.}
    \label{delta_redshift}
\end{figure*}

Also, we estimate the outlier fraction defined as,  $\frac{|z_{\mathrm{ph}} - z_{\mathrm{spec}}|}{(1+z_{\mathrm{spec}})} > 0.2$ \citep{Duncan2018}, to asses the precision of photometric redshifts. The outlier fraction makes up $\sim$ 13\% of the total samples in both cases.  In Fig. \ref{delta_redshift}, we show the variation of $\frac{|z_{\mathrm{ph}} - z_{\mathrm{spec}}|}{(1+z_{\mathrm{spec}})}$ as a function of spectroscopic redshifts.  We find that for both cases, 400 MHz and 612 MHz catalogue, the estimate of photometric redshifts have some outliers with respect to spectroscopic redshifts. We find that the standard deviation of outlier fraction is $\sim 0.6$ for both catalogues. However, neglecting those  outliers from the analysis does not change our statistical estimation of angular and spatial correlation functions. The reason behind large deviation of photometric redshift with respect to spectroscopic redshift for some sources is unknown. In future work we will use more updated redshift catalogue for this particular field. 

\subsection{Classification of sources}
\label{sec_classify}
In this work we  classify the catalogued sources  at 400 MHz  and 612 MHz with redshift information as AGNs or SFGs using their radio luminosity as a sole indicator.
 \citet{McAlpine2013} show that the luminosity function evolves  as $(1+\textit{z})^{2.5}$ for SFGs but for AGN powered galaxies as $(1+\textit{z})^{1.2}$. Further, \citet{Magliocchetti2014,Magliocchetti2016,Magliocchetti2017} showed that AGN powered galaxies become dominant beyond the radio power $P_{\mathrm{cross}} (z)$, which scales with redshift $z$ as : 
\begin{equation}
    \mathrm{log_{10}}P_{\mathrm{cross}} = \mathrm{log_{10}}P_{0,\mathrm{cross}} + \textit{z} ,
    \label{eqn_classify}
\end{equation}
to $z \sim 1.8$,  where radio power $P$ is in $\mathrm{W Hz^{-1}sr^{-1}}$. In the local Universe $P_{0,\mathrm{cross}} = 10^{21.7}$ [$\mathrm{W Hz^{-1}sr^{-1}}$] and nearly coincides with  the
break in the radio luminosity function of star-forming galaxies \citep{Magliocchetti2002}.  Beyond that the luminosity function of SFGs decreases rapidly and their population is drastically reduced. The  possibility of contamination between the AGNs and SFGs  using a luminosity based selection  is quite low \citep{Magliocchetti2017}. 

The radio luminosities of sources in EN1 field with  redshift information can be calculated from their flux using \citep{Magliocchetti2014}: 
\begin{equation}
    P_{\mathrm{1.4GHz}} = S_{\mathrm{1.4GHz}} D^{2} (1+\textit{z})^{3+\alpha} ,
\end{equation}
where $P_{\mathrm{1.4GHz}}$ is in [$\mathrm{W Hz^{-1}sr^{-1}}$] units, $\textit{D}$ is the angular diameter distance and $\alpha$ is the spectral index of radio emission ($S_{\nu} \propto \nu^{-\alpha}$). Note that, we do not have pre-measured spectral index values for all the sources in our catalogue.  \citet{Chakraborty2019B} compared the flux densities of matched sources in EN1 field with high frequency catalogues and found a median spectral index of 0.8. Since we are probing the faint sources here, the probability of finding a large number of bright, flat spectrum AGNs is low \citep{Magliocchetti2017}. We consider the spectral index for all the sources as 0.8 to estimate their radio luminosities.

\begin{figure*}
    \centering
   
\begin{tabular}{c|c|c|c}
       \includegraphics[width=\columnwidth,height=2.5in]{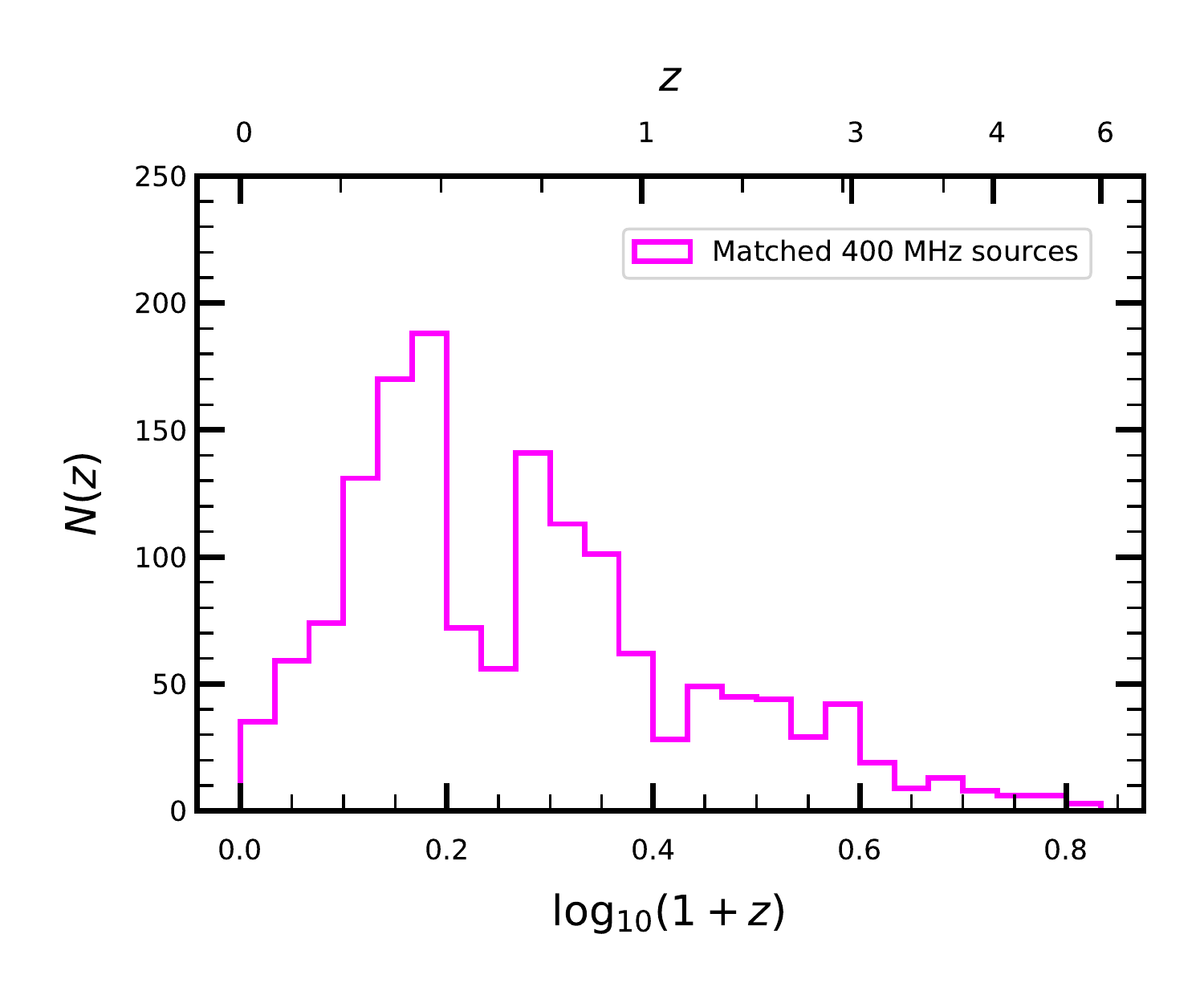} & 
        \includegraphics[width=\columnwidth,height=2.5in]{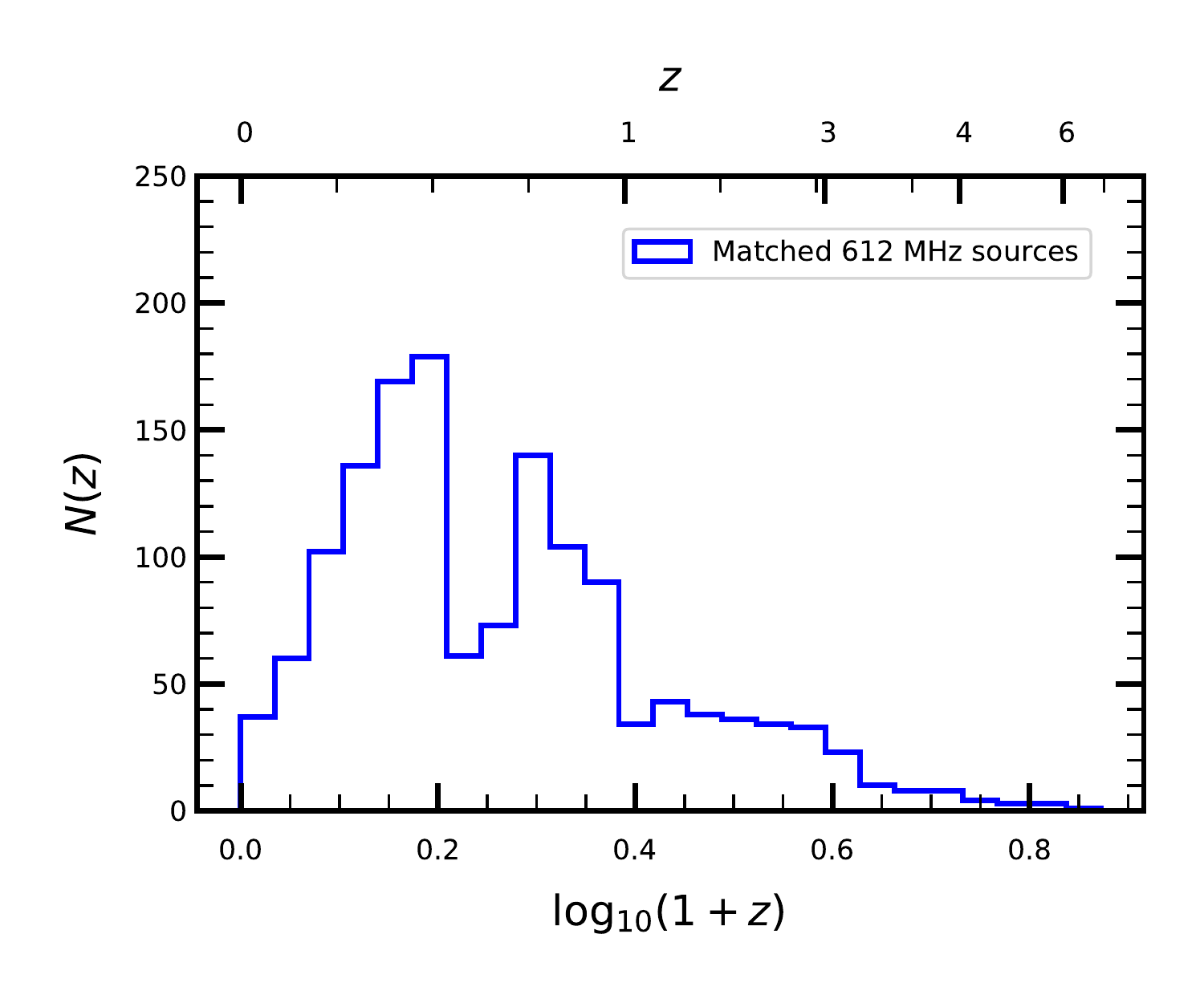} \\ 
       \includegraphics[width=\columnwidth,height=2.5in]{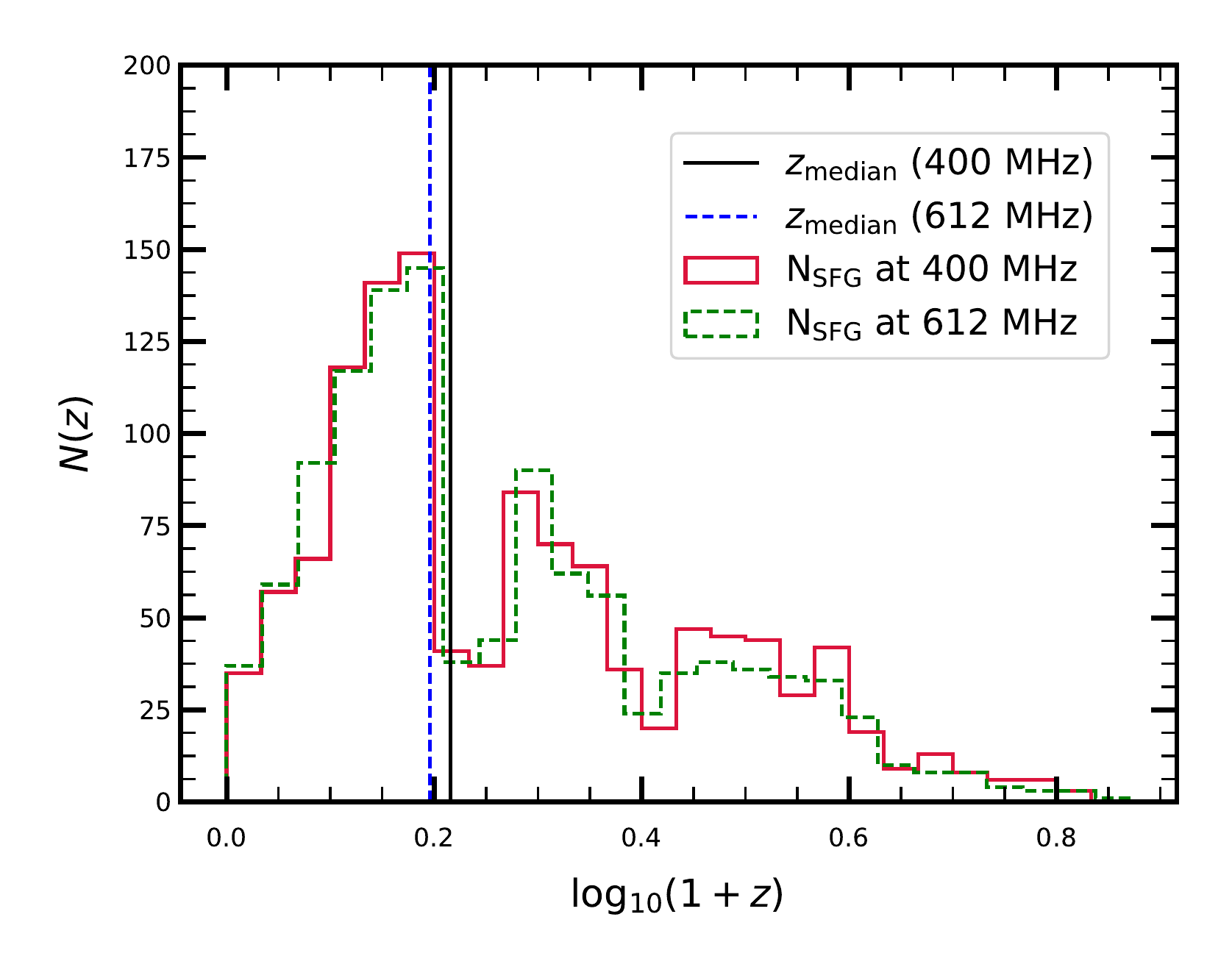} &
       \includegraphics[width=\columnwidth,height=2.5in]{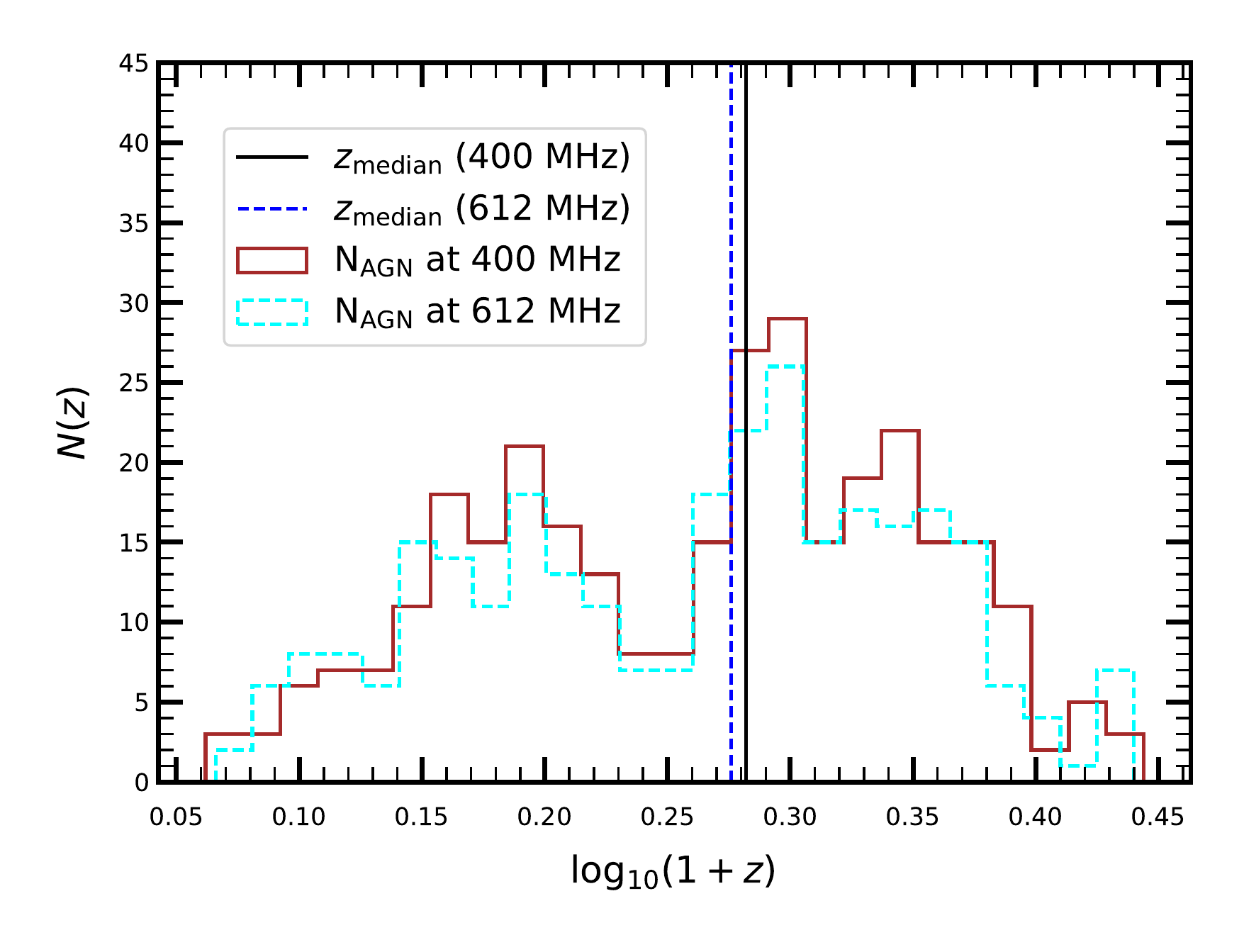}    
      
\end{tabular}      
         
    \caption{The redshift distribution (N($z$))of matched sources with the SWIRE and SDSS catalogues for 400 MHz (top left) and  612 MHz (top right) catalogue respectively. The redshift distribution of SFGs (bottom left) and AGNs (bottom right) at 400 MHz and 612 MHz are also shown. {\bf Here the redshift of sources includes both photometric and spectroscopic redshifts.} The black and blue  dashed lines in the bottom panel are the median values of the  redshift distribution at the corresponding frequency.}
    \label{AGN_SFG_distribution}
\end{figure*}

We classify the catalogued sources  with redshifts $z \leq 1.8$ as AGNs if their luminosity is greater than the threshold given in  Eqn. \ref{eqn_classify} and SFGs otherwise. For higher redshifts, we fix $P_{0,\mathrm{cross}} = 10^{23.5}$ [$\mathrm{W Hz^{-1}sr^{-1}}$] \citep{McAlpine2013}. Following this procedure, out of 1503 sources at 400 MHz (with redshift information), 314 (20.9\%)  and 1189 (79.1\%) sources has been identified as AGNs and SFGs respectively.  Whereas, at 612 MHz, out of 1429 sources (with redshifts) 290 (20\%) and 1139 (79\%) sources has been identified as AGNs and SFGs respectively.  The redshift distributions of AGNs and SFGs  are shown in Fig. \ref{AGN_SFG_distribution}. The median value of redshift  for SFGs is 0.64 and that for the AGNs is 0.91 at 400 MHz. For 612 MHz catalogue the median redshift for SFGs is 0.57 and for AGNs it is 0.85.  

\section{Estimation of the correlation functions}

\subsection{The angular correlation function} 
\label{sec_angular}
 The angular two-point correlation function, $w(\theta)$, quantifies  clustering of sources on angular scale in sky-plane. 
A detailed discussion on different estimators of $w(\theta)$ can be found  in literature (see: \citealt{Kerscher2000}). In this work, we  use the  LS estimator proposed by \citet{Landy1993}:
\begin{equation}
    w(\theta) = \frac{DD(\theta) - 2 DR(\theta) + RR(\theta)}{RR(\theta)}.
    \label{LS_estimator}
\end{equation}
Here, $DD(\theta)$  and $RR(\theta)$ are the  average count of pair of objects separated by an angle $\theta$ in original catalogue and in a random catalogue, respectively. $RR(\theta)$ is calculated from a large simulated catalogue constructed by distributing sources randomly across the same survey area as the real data.  This estimator also includes the count of cross-pair separation, $DR(\theta)$, between real and random catalogue, which effectively reduces the large scale uncertainty in source density \citep{Cress1996,Overzier2003}.  


The error  in $w(\theta)$ is being calculated using  `bootstrap' re-sampling method  \citep{Ling1986}. 
Here, we have generated $100$ `bootstrap' to quote  uncertainty in $w(\theta)$ measurement.

    

\begin{table*}
\caption{Clustering parameters from this study and previous obsevations}
\label{clustering_paramateres}
\scalebox{1.0}{
\begin{tabular}{|c|c|c|c|c|c|c|c|}
\hline
\hline
Observation & $\nu$ (MHz) &  $S_{\mathrm{limit,\nu}}^{\dagger}$ (mJy) & log$_{10}$(A) & $\gamma$ & $S_{\mathrm{cut,400MHz}}$ (mJy) & $S_{\mathrm{cut,612MHz}}$ (mJy) & Ref \\
\hline 

FIRST & 1400 & 1 & $-2.30_{-0.9}^{+0.7}$ & $1.82 \pm 0.02$ & 2.72 & 1.93 & \citet{Lindsay2014} \\
\hline 
COSMOS & 3000 & 0.013 & $-2.83_{-0.1}^{+0.1}$ & 1.80 & 0.065 & 0.046 & \citet{Hale2018} \\
\hline
TGSS & 150 & 50 & $-2.11_{-0.3}^{+0.3}$ & 1.82 & 22 & 16.0  & \citet{Rana2019}\\

\hline
XMM-LSS & 144 & 1.4 & $-2.08_{-0.04}^{+0.05}$ & 1.80 & 0.61 & 0.44 & \citet{Hale2019}\\
\hline
EN1 & 400 & 0.100 & $-2.03_{-0.08}^{+0.10}$ & $1.75 \pm 0.06$ & 0.100 & 0.071 & This work \\
\hline
EN1 & 612 & 0.050 & $-2.22_{-0.15}^{+0.12}$ & $1.81 \pm 0.06$ & 0.070 & 0.050 & This work \\

\hline
\end{tabular}}
\flushleft{${\dagger}$ $S_{\mathrm{limit,\nu}}$ is the flux density limit of the corresponding catalogue at observed frequency ($\nu$). \\
 }
\end{table*}

\subsection{Random Catalogue}

The noise is not uniform across the map of the EN1 field. This results in systematic non detection of fainter sources at the regions of the field of view with higher noise and hence introduces a bias in the estimate of the angular two point correlation function. The effect of inhomogeneous noise in the image needs to be incorporated in the generation of random catalogue to avoid this bias. We use the following procedure to generate random catalogue in order to estimate the angular correlation function unbiasedly.

We use P{\tiny Y}BDSF to  estimate the  noise map for the EN1 field. Assuming that the radio sources follow the flux distribution of $S^{3}$ catalogue \citep{Wilman2008}, we randomly choose $1000$ sources from the $S^{3}$ catalogue with flux densities  in a given range. We simulate a radio map with compact sources with the above scaled flux densities distributed randomly in the EN1 field. We generate a mock sky map with these sources  and add the  above noise map. We use P{\tiny Y}BDSF with same parameters as that used to estimate  the EN1 catalogue (see Sec.4 in \citealt{Chakraborty2019B})  to construct the random catalogue from this map. This gives us one realization of the random catalogue that can be used to estimate the angular correlation function unbiasedly. We use $100$ statistically independent realizations of such random catalogues  to reduce the statistical uncertainty associated with it.

Since the noise map of the  400 MHz and 612 MHz data are different we generated  separate random catalogues for them following the  procedure described above. For the $400$ MHz catalogue we draw sources with flux densities in the range between $75 \mu$Jy to $1$ Jy from the $S^{3}$ catalogue, whereas for the 612 MHz catalogue the sources were selected within flux density range of  $40 \mu$Jy to $1$ Jy, where lower flux density threshold  corresponds to the 5$\sigma$ limits of the corresponding maps.

\begin{figure*}
    \centering
    \begin{tabular}{c|c}
     \includegraphics[width=\columnwidth,height=3in]{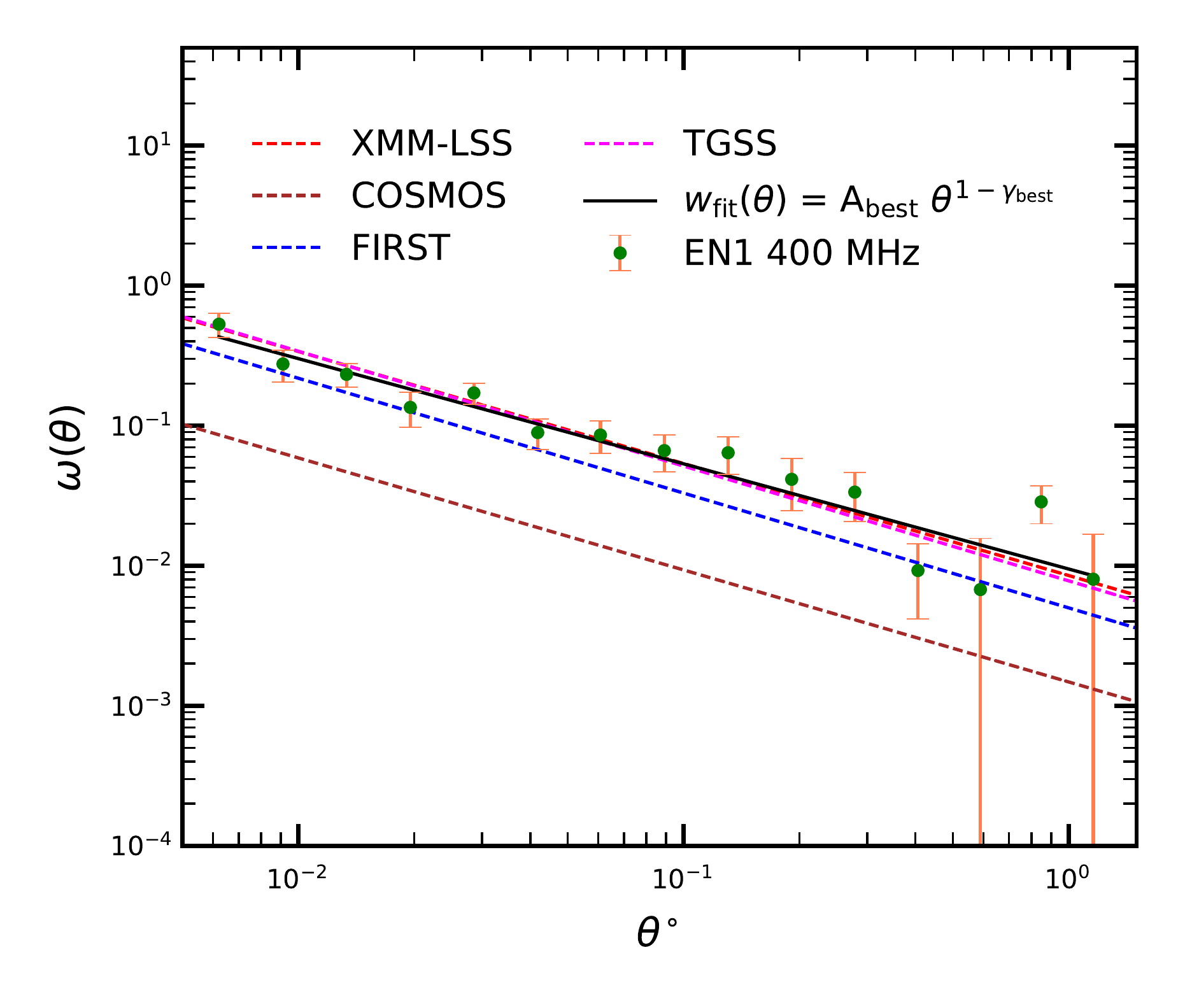} &
     \includegraphics[width=\columnwidth,height=3in]{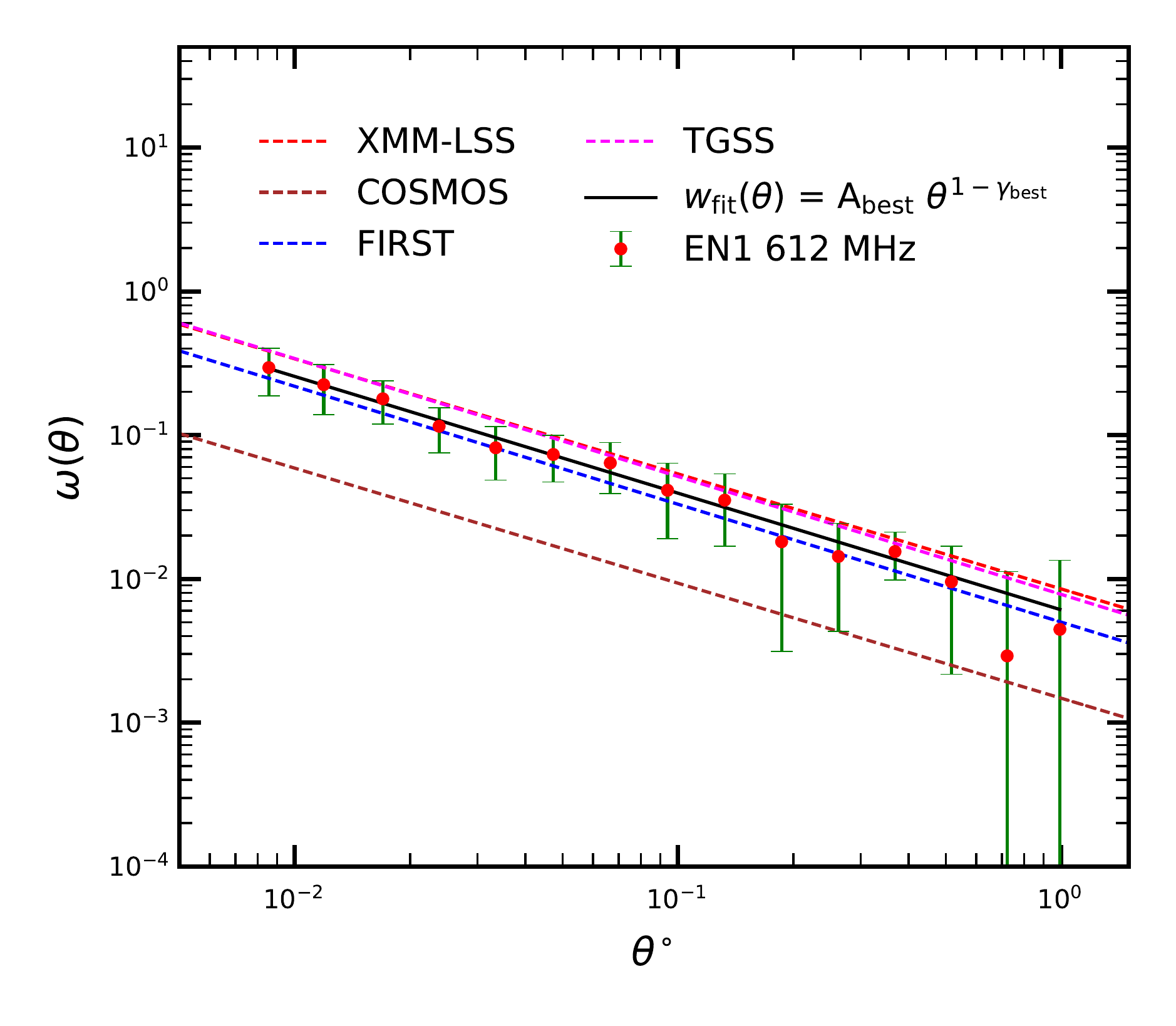} 
    \end{tabular}
    \caption{ Angular correlation function of sources in EN1 using 400 MHz (left) and 612 MHz (right) catalogue. The black line is  the best fitted power law to $w(\theta)$. We also compare our findings with previous observations of XMM-LSS \citep{Hale2019}, FIRST \citep{Lindsay2014}, COSMOS \citet{Hale2018} and TGSS \citep{Rana2019}.  }
    \label{w_theta_radio}
\end{figure*}

\subsection{Results of clustering pattern of compact sources at 400 MHz and 612 MHz}
We use the publicly available code TreeCorr \footnote{\url{https://github.com/rmjarvis/TreeCorr}} \citep{Jarvis2004}  to find the angular correlation function of  the sources in the 400 MHz  catalogue in 15 equal  logarithmically spaced bins between $ \theta \sim 18\arcsec - 1.5^{\circ}$.  The lower limit for the correlation is chosen to be 4 times the synthesized beam (PSF). Note that the synthesized beam at 612 MHz is higher than 400 MHz. This is mainly because of unavailability of large baselines at 612 MHz due to calibration and editing. The upper limit is restricted to the $\sim$1.5 times the HPBW (primary beam) at respective frequencies. The left panel of Fig. \ref{w_theta_radio},  shows  the estimated value of $w(\theta)$ at $400$  MHz as a function of $\theta$ with (orange) filled circles. The error bars are estimated by `bootstrap' re-sampling method as described earlier. We fit $w(\theta)$ with a power law of the form $w(\theta) = A \theta^{1-\gamma}$. We use Markov Chain Monte Carlo (MCMC) simulation for this fitting and parameter estimation. We use Metropolis-Hastings algorithm to simulate $10^{6}$ data points in the $A -\gamma$ parameter space and use $\chi^{2}$ to estimate the most likely values of the  parameters.  The best fitted values are $\mathrm{log(A)} = -2.03^{+0.08}_{-0.10}$ and $\gamma = 1.75 \pm 0.06$, where the error bars are  $1-\sigma$ error bars corresponding to 68.3\% MCMC points.  
\begin{figure}
    \centering
    \includegraphics[width=\columnwidth,height=3in]{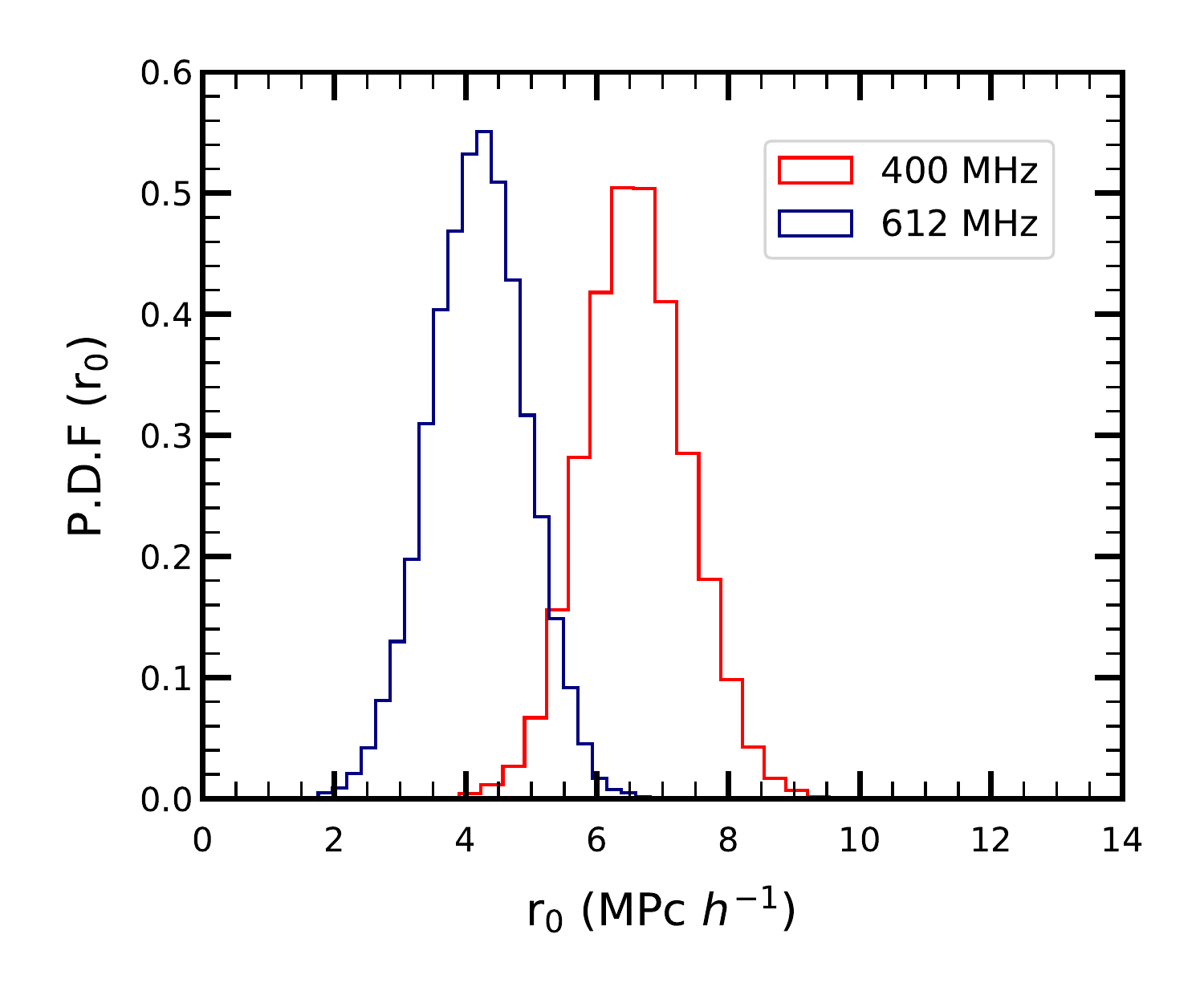} 
    \caption{The probability distribution function (PDF) of the spatial clustering length ($r_{0}$) for 400 MHz (red) and 612 MHz (blue) catalogues.}
    \label{spatial_scale}
\end{figure}

\begin{figure*}
    \centering
    \begin{tabular}{c|c}
          \includegraphics[width=\columnwidth,height=2.9in]{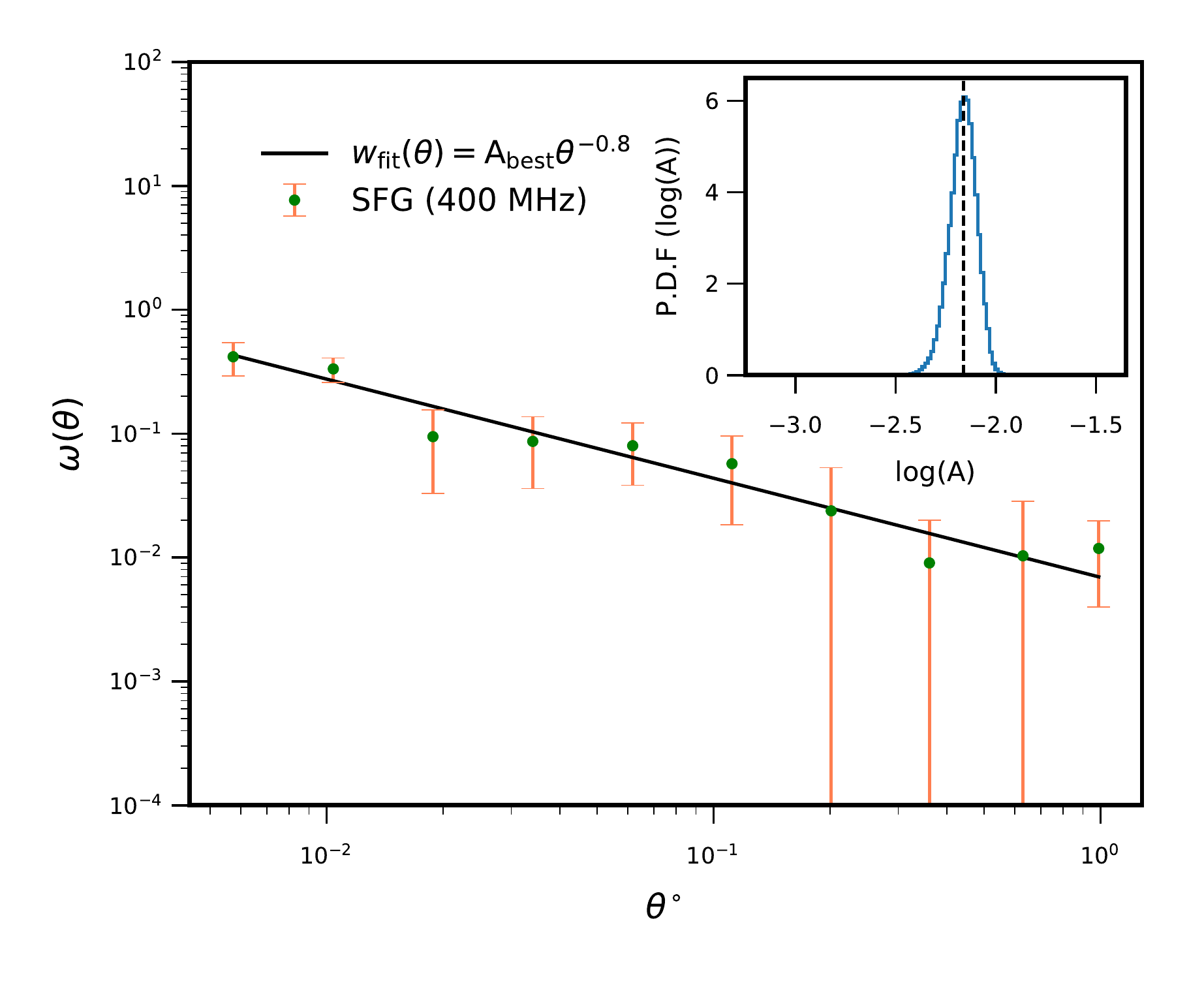} & 
          \includegraphics[width=\columnwidth]{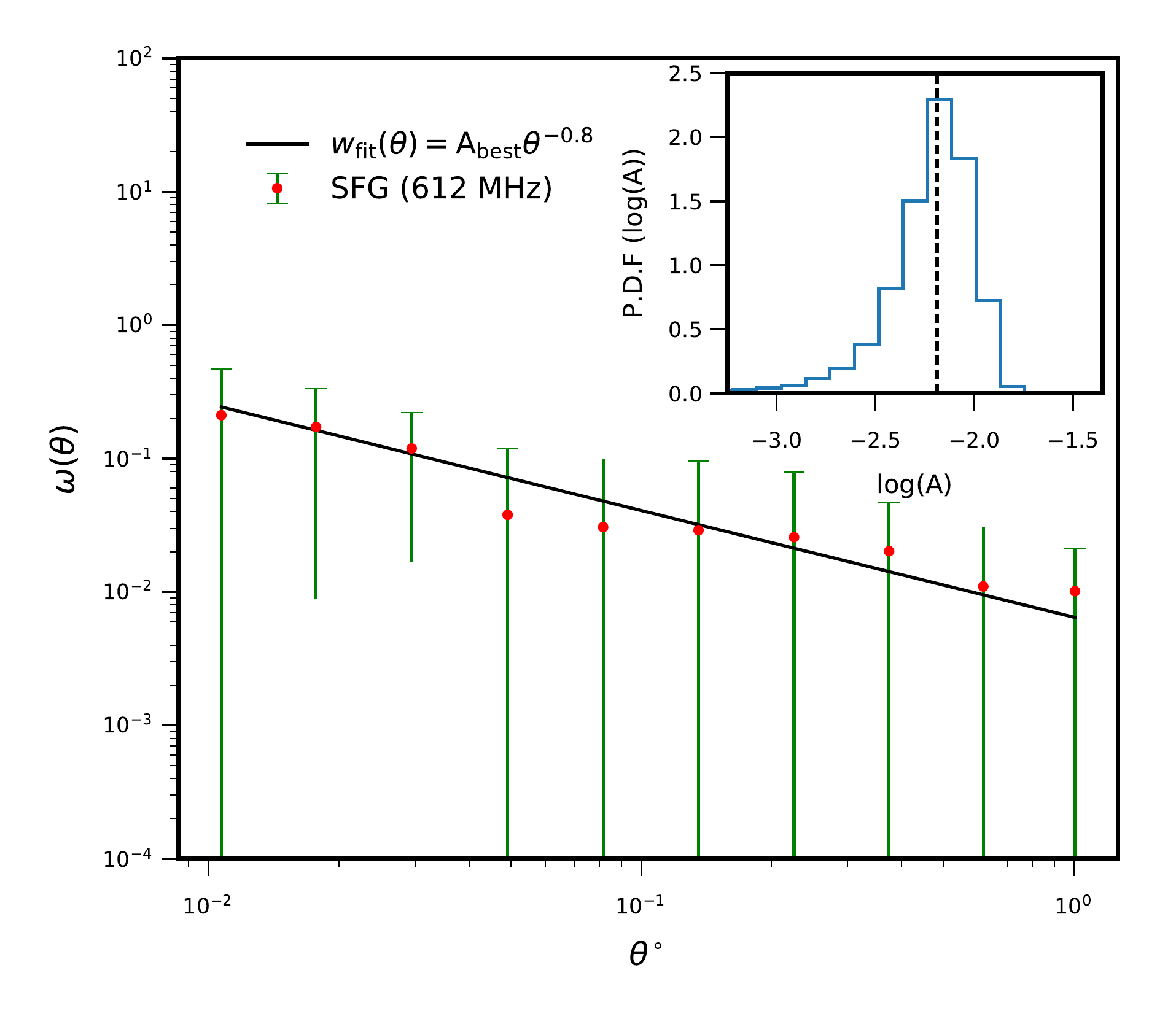} \\
         \includegraphics[width=\columnwidth,height=3in]{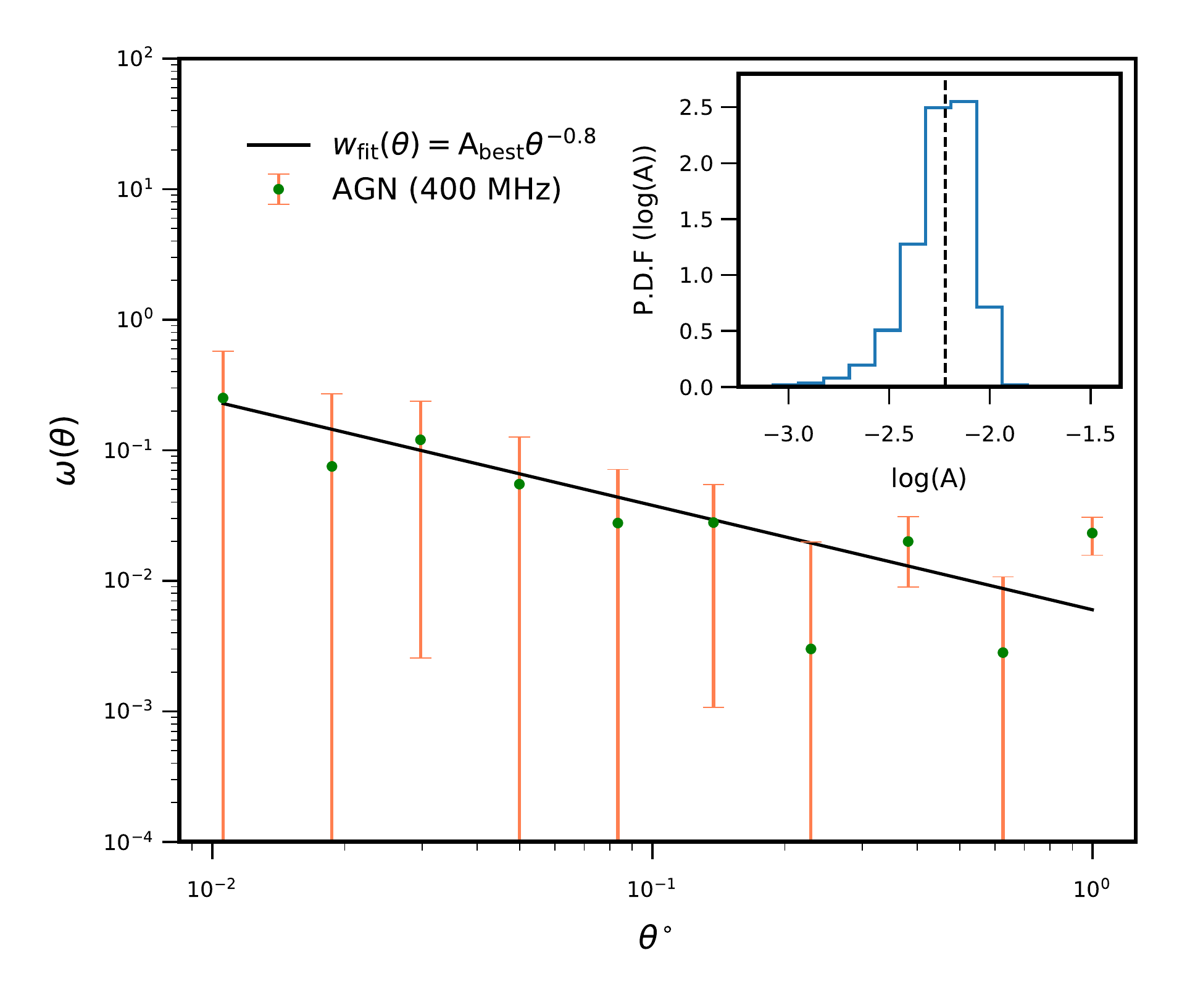} &
         \includegraphics[width=\columnwidth]{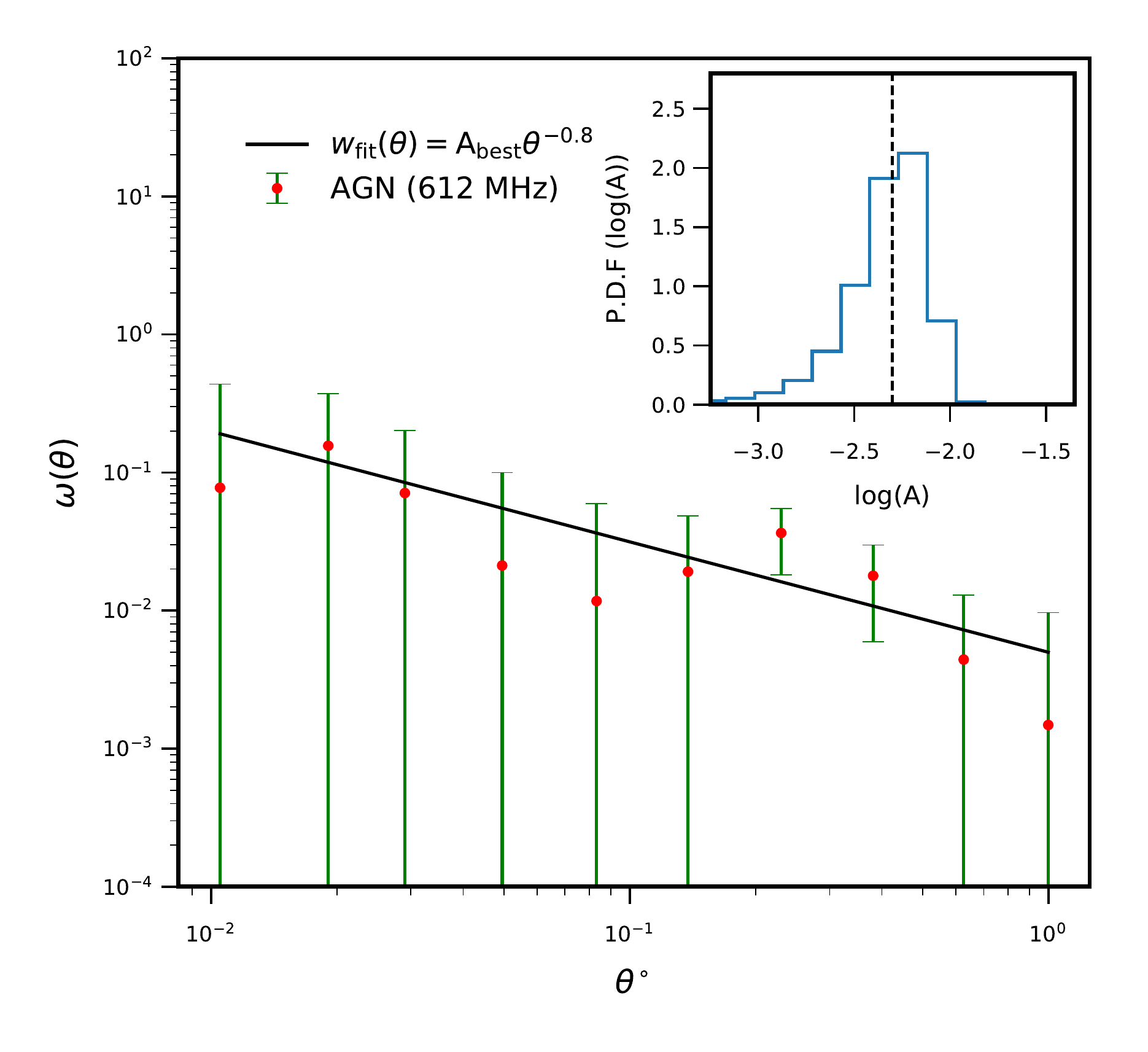} \\
    \end{tabular}
        
    \caption{The angular correlation of SFGs (upper panel) and AGNs (lower panel) along with the probability distribution function of amplitude of clustering (insets) when the power-law index $\gamma$ is fixed. The left column is for 400 MHz catalogue and right column is for 612 MHz catalogue.}
    \label{clustering_AGN_SFG}
\end{figure*}

The right panel of Fig. \ref{w_theta_radio} shows the angular correlation function at 612 MHz as estimated in  15 equally spaced logarithmic bins between $ \theta \sim 25\arcsec - 1.2^{\circ}$.  The best fit values of the parameters are  $\mathrm{log(A)} = -2.22^{+0.12}_{-0.15}$ and $\gamma = 1.81 \pm 0.06$. 

\subsection{Comparison to previous observations}
The results of angular clustering from various studies along with the findings in this paper is summarized in Table~\ref{clustering_paramateres}. Column (7) and (8) in the Table~\ref{clustering_paramateres} gives the  projected flux density cut offs at 400 and 612 MHz (assuming a spectral index of $0.8$; $S_{\nu} \propto \nu^{-\alpha}$ ) for  different catalogues used to calculate the angular correlation function. 
The best fit estimate of the parameter $\gamma$ is found to be consistent with all the previous observations at different frequency bands as well as with its theoretical prediction \citep{Peebles1980}. 

Given similar types of sources in a population, the less luminous sources are expected to be found in weakly clustered less massive haloes. This results in measurement of lower value of the amplitude of the correlation function for deeper catalogues.  We believe this the reason of higher clustering amplitude in the 400 MHz catalogue as compared to the 612 MHz catalogue. 

\begin{table*}
\caption{Spatial clustering and bias from this study and previous obsevations}
\label{spatial_clustering_paramateres}
\scalebox{1.0}{
\begin{tabular}{|c|c|c|c|c|c|c|c|}
\hline
\hline
Observation & Source type & Median redshift &  log$_{10}$(A) & $r_{0}$ & b($z_{\mathrm{median}}$) & Ref \\
\hline 
         & AGN & 0.70 &  $-2.30^{+0.1}_{-0.1}$ &$6.9^{+0.60}_{-0.70}$ & $2.1^{+0.2}_{-0.2}$ & \citet{Hale2018}\\
COSMOS   & AGN & 1.24 &  $-2.60^{+0.1}_{-0.1}$ &$9.6^{+0.70}_{-0.70}$ & $3.6^{+0.2}_{-0.2}$ & \citet{Hale2018}\\
(3GHz)   & AGN & 1.77 &  $-2.60^{+0.1}_{-0.1}$ & $7.3^{+0.90}_{-0.90}$ & $3.5^{+0.4}_{-0.4}$ & \citet{Hale2018}\\
         & SFG & 0.62 &  $-2.60^{+0.1}_{-0.1}$ & $5.0^{+0.50}_{-0.60}$ & $1.5^{+0.1}_{-0.2}$ & \citet{Hale2018}\\
         & SFG & 1.07 &  $-2.90^{+0.1}_{-0.1}$ & $6.1^{+0.60}_{-0.70}$ & $2.3^{+0.2}_{-0.2}$ & \citet{Hale2018}\\
\hline
VLA-COSMOS & AGN & 1.25 &  $-2.79^{+0.1}_{-0.1}$ &$7.84^{+1.75}_{-2.31}$ & - & \citet{Magliocchetti2017}\\
 (1.4GHz)  & SFG & 0.50 &  $-2.36^{+0.3}_{-0.3}$ &$5.46^{+1.12}_{-2.10}$ & - & \citet{Magliocchetti2017}\\

\hline
ELAIS N1  &  AGN & 0.91 &  $-2.22^{+0.16}_{-0.16}$ &$7.30^{+1.4}_{-1.2}$ & $3.17^{+0.5}_{-0.4}$ & This work \\
(400 MHz) &  SFG & 0.64 & $-2.16^{+0.05}_{-0.06}$ &$4.62^{+0.39}_{-0.40}$ & $1.65^{+0.14}_{-0.14}$ & This work \\
\hline
ELAIS N1 & AGN & 0.85 &  $-2.30^{+0.02}_{-0.03}$ &$6.0^{+1.5}_{-1.3}$ & $2.6^{+0.6}_{-0.5}$ & This work \\
 (612 MHz) &  SFG & 0.57 & $-2.18^{+0.01}_{-0.02}$ &$4.16^{+0.7}_{-0.8}$ & $1.59^{+0.2}_{-0.2}$ & This work \\

\hline
\end{tabular}}
\end{table*}
\citet{Hale2018} measured angular two-point correlation function of radio sources in Cosmological Evolution Survey (COSMOS) field at 3 GHz \citep{Smolcic2017A,Smolcic2017B}. \citet{Hale2018} reports estimation of clustering amplitude of all sources as $\mathrm{log(A)} = -2.83_{-0.1}^{+0.1}$  in this field using a $\sim$ 13 $\mu$Jy beam$^{-1}$ flux density limit. The amplitude for this deep survey suggests less clustering of the fainter sources as expected.
However, it needs to be noted that at $3$ GHz, COSMOS is also likely to look at different population of sources and a direct comparison of the clustering observed in it with the results presented here may not be straight forward.

\citet{Hale2019} reports a clustering amplitude  of $\mathrm{log(A)} = -2.08_{-0.05}^{+0.04}$ in the XMM-LSS field at 144 MHz. \citet{Rana2019} investigate the clustering  for sources in the TIFR GMRT Sky Survey  at 150 MHz (TGSS-ADR1) \citep{Intema2017} over a large fraction of  sky and  show  variation of clustering amplitude as a function of  different flux density cut offs.  Clustering properties of sources in the Faint Images of the Radio Sky at Twenty-cm (FIRST) survey at 1400 MHz \citep{Becker1995} is given in  \citet{Lindsay2014}. The amplitude of the clustering in these surveys are statistically consistent with what we report in this work. 


\section{Estimation of spatial Correlation function from the redshift distribution of sources}
\label{sec_spatial}


 Spatial clustering of sources is quantified by the two-point correlation function $\xi(r)$.  With known  angular clustering $w(\theta)$ of the sources and  the redshift distribution, $N(z)$, it is possible to use the Limber inversion \citep{Limber1953}  method to estimate$\xi(r)$. We briefly discuss this technique here.

Due to gravitational clustering, the spatial clustering of sources changes with redshift. We assume a epoch-dependent power-law  \citep{Overzier2003}  spatial correlation function as: 
\begin{equation}
    \xi(r,\textit{z}) = (r_{0}/r)^{\gamma} (1+\textit{z})^{\gamma - (3+\epsilon)},
\end{equation}
where r is in comoving units,  and $\epsilon$ quantifies different clustering models \citep{Overzier2003}. The parameter $r_0$ gives the spatial clustering length at $z=0$. In this work we use comoving clustering model with $\epsilon = \gamma -3$.  In this model, the correlation function remains unchanged in comoving coordinates and the cluster has fixed comoving size. The spatial correlation length in comoving coordinates can be estimated as  \citep{Peebles1980}: 
\begin{equation}
    A = r_{0}^{\gamma} H_{\gamma} \Big(H_{0}/c\Big) \frac{\int_{0}^{\infty} N^{2}(\textit{z})(1+\textit{z})^{\gamma - (3+\epsilon)} \chi^{1-\gamma}(\textit{z})E(\textit{z})d\textit{z}}{\big[\int_{0}^{\infty} N(\textit{z}) d\textit{z}\big]^{2}},
    \label{limber_inversion}
\end{equation}
where,  $H_{\gamma} = \frac{\Gamma(\frac{1}{2}) \Gamma(\frac{\gamma -1}{2})}{\Gamma(\frac{\gamma}{2})}$, 
   $ E(\textit{z}) = [\Omega_{m,0}(1+\textit{z})^{3} + \Omega_{k,0}(1+\textit{z})^{2} + \Omega_{\Lambda,0}]^{1/2}$,
  $N(\textit{z})$ is the redshift distribution of sources and $\chi({\textit{z}})$ is the  line-of-sight comoving distance. We use Eqn. \ref{limber_inversion} to estimate the spatial correlation length, $r_{0}$, from the estimated  clustering amplitude  of angular correlation function A, slope of correlation function $\gamma$ and redshift distribution $N(z)$ \citep{Lindsay2014}.

The estimated value of the parameter $\gamma$ across different surveys as mentioned in Table \ref{clustering_paramateres} agrees with the theoretical predicted value of $1.8$ \citep{Peebles1980}. We use $\gamma = 1.8$, the MCMC distribution of the parameter $A$ as given in section \ref{sec_angular} and the redshift distribution of sources as given in section \ref{redshift_info} to estimate $r_{0}$. Fig. \ref{spatial_scale} shows the probability density function for $r_{0}$ in 400 and 612 MHz. The median value of $r_{0}$ along with 16th and 84th percentile errors are $6.58 ^{+0.75}_{-0.83}$ at a median redshift of $\textit{z} = 0.803$ for 400 MHz observation. For 612 MHz observation, the clustering length is  $4.19 ^{+0.73}_{-0.74}$ at a median redshift of $\textit{z} = 0.70$. The lesser value of the clustering length at the 612 MHz is a direct consequence of the lesser clustering of the fainter sources detected at this band.

   

\subsection{Bias} 
\label{sec_bias}
The relation between clustering properties of luminous sources and underlying dark matter is described by the bias parameter. The scale independent linear bias parameter $b(z)$ is defined as  the ratio of the galaxy to the dark matter spatial correlation functions \citep{Kaiser1984,Bardeen1986,Peacock2000}.
For a standard cosmological model the bias parameter can be calculated from the spatial clustering length $r_0$  using \citep{Lindsay2014,Hale2018}:
\begin{equation}
    b(\textit{z}) = \Big(\frac{r_{0}(\textit{z})}{8}\Big)^{\gamma/2} \frac{J_{2}^{1/2}}{\sigma_{8}D(z)/D(0)},
\end{equation}
were, $J_{2} = 72/[(3-\gamma)(4-\gamma)(6-\gamma)2^{\gamma}]$, $D(z)$ is the linear growth factor  \citep{Hamilton2001} and  $\sigma^{2}_{8}$ is the dark matter density variance in a comoving sphere of radius 8 Mpc $h^{-1}$. 
In this work,  we have estimated the bias at the median redshift  using the median value of $r_{0}$ corresponding to that redshift along with 16th and 84th percentile errors. We find that at 400 MHz $b_{z} = 2.74^{+0.27}_{-0.30}$ at median redshift of $\textit{z} = 0.803$ and at  612 MHz  $b_{z} = 1.73^{+0.27}_{-0.28}$ at a median redshift of $0.7$.


\section{Clustering of SFGs and AGNs}
\label{sec_AGN_SFG}
 
In this section, we discuss results of our estimation of   the clustering properties and corresponding bias  of two different samples contributing to the total source counts, the AGNs and SFGs, at 400 MHz and 612 MHz. The evolution of clustering for AGNs and SFGs are expected to be different.  Given that the median redshift of the two samples are different in the two observing bands, we expect to trace the hint of redshift evolution of different sources here. Following the similar procedure as described earlier, we estimate angular correlation of AGNs and SFGs and find the best fit value for the amplitude of correlation.   The low density of these two samples makes it statistically inappropriate to perform a two parameter fit to the measured correlation function.  In view of this, we fix the $\gamma$ to theoretically predicted value of 1.8 and  fit for the amplitude $A$ only. Note that the reduced chi square value giving the goodness of the fit for both the SFG ad AGNs are $\sim$0.1 suggesting that these results has to be taken with caution. However, this analysis allow us to check the consistency of correlation function of SFG's and AGN's with previous findings at radio as well as at other wavelengths. In Fig. \ref{clustering_AGN_SFG}, the estimated $w(\theta)$ and the best fit model for SFGs  and AGNs are shown. These results are summarized in Table \ref{spatial_clustering_paramateres}. 
 The spatial clustering length for SFGs at a median redshift $z_{\mathrm{median}} = 0.64$ is found to be $r_{0}=4.62^{+0.39}_{-0.40}\  h^{-1}$ Mpc at 400 MHz, which decreases to  $r_0 = 4.16^{+0.7}_{-0.8}\  h^{-1}$ Mpc at the median redshift of $0.57$ at 612 MHz. The corresponding bias for SFGs are $b_{\textit{z}}$ =  $1.65^{+0.14}_{-0.14}$ and $b_{\textit{z}}$ = $1.59^{+0.2}_{-0.2}$ at 400 MHz and 612 MHz respectively. For the AGN population the clustering length is relatively higher in general. At $z=0.91$ the clustering length for the AGNs is $r_0 = 7.3^{+1.4}_{-1.2}\  h^{-1}$ Mpc at 400 MHz which decreases to $r_0 = 6.0^{+1.5}_{-1.3} \  h^{-1}$ Mpc at a redshift of $z=0.85$ at 612 MHz. The bias for AGNs are $b_{\textit{z}} = 3.17^{+0.5}_{-0.4}$ and $b_{\textit{z}} = 2.6^{+0.6}_{-0.5}$ for 400 and 612 MHz catalogue respectively. We find that  there is a hint of decrease of the bias parameter  for both populations at higher frequency as the median redshift of these populations decreases. This indicates that we are most likely tracing the fainter less massive haloes at higher frequency.  The variation of bias parameter for SFGs and AGNs along with some previous findings are shown in the Fig \ref{overplot_bz}. We also show in Fig. \ref{overplot_bz},  the assumed redshift evolution of bias parameters of different kind of sources in $S^3$ simulation.  The fact that the estimated bias for AGN is higher compared to that for the SFGs is consistent with results from $S^3$ simulation. The large bias parameter  indicates that the AGNs are strong biased  tracers of dark matter halos than SFGs.
 Note that, due to  uncertainties associated with measurement of redshift, clustering length and bias parameter and owing to the values of almost similar redshifts for different frequency bands, we have not observed much redshift evolution of these different populations here.
\begin{figure*}
    \centering
    \begin{tabular}{c|c}
     \includegraphics[width=\columnwidth,height=3in]{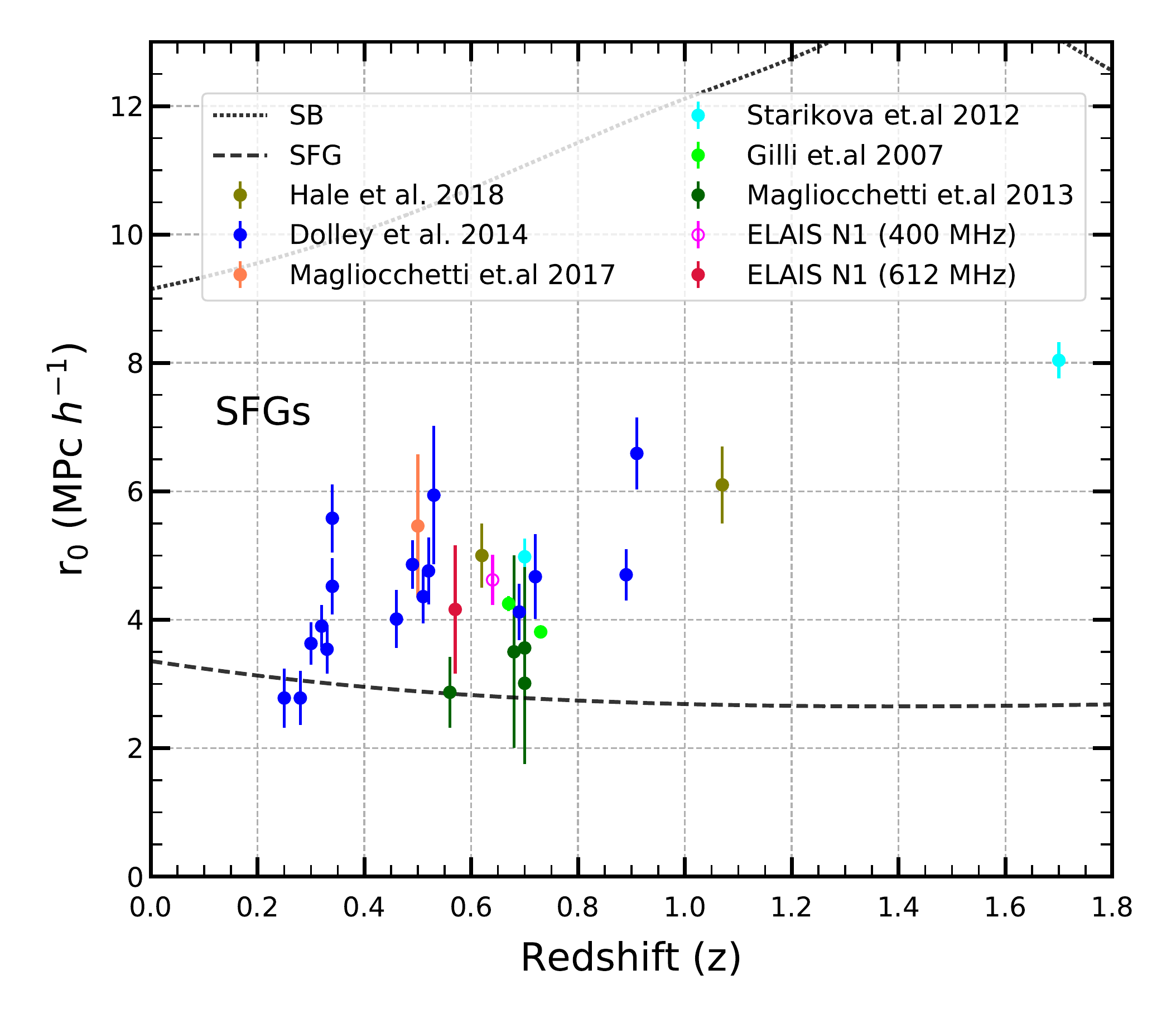} &
     \includegraphics[width=\columnwidth,height=3in]{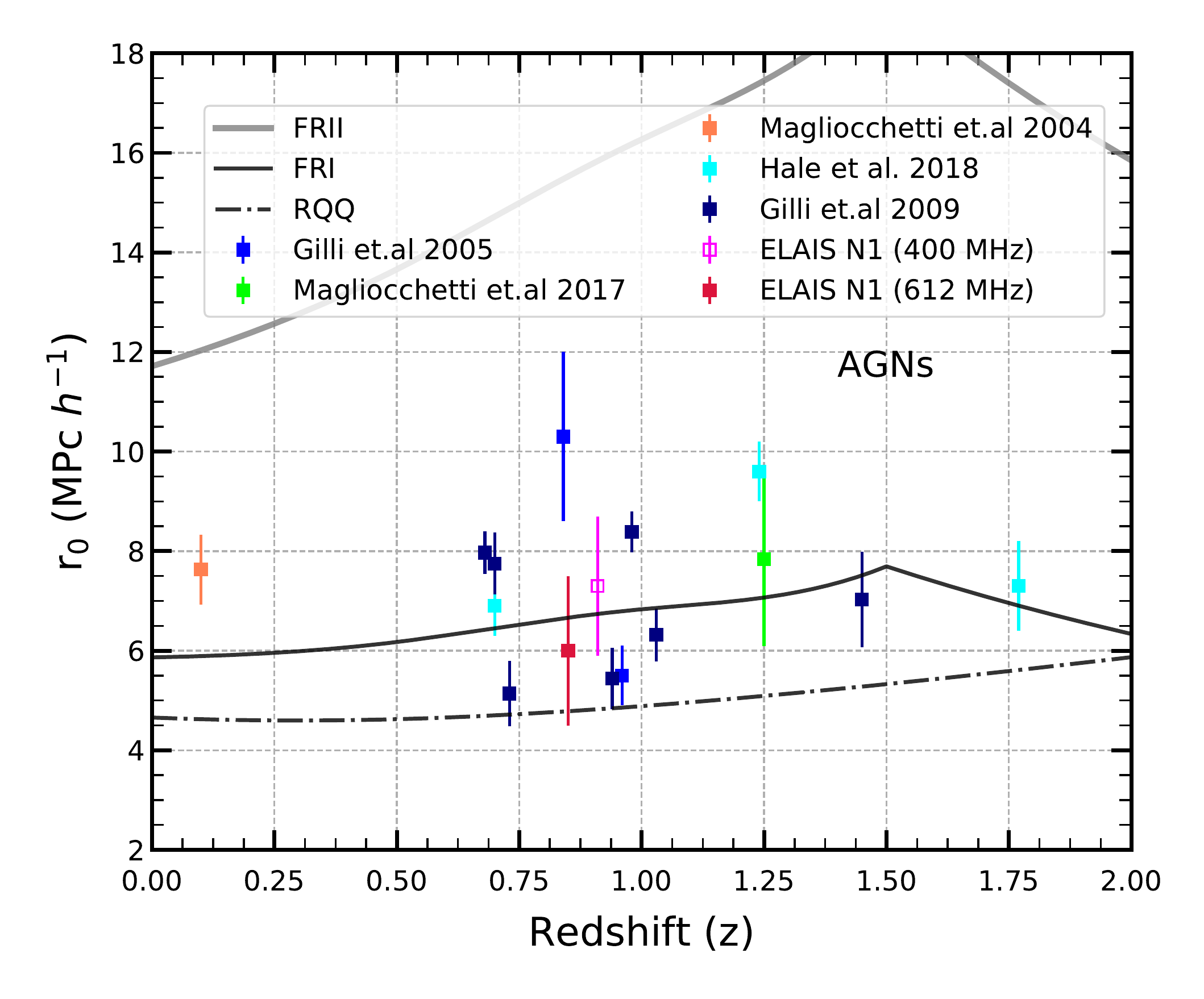} 

    \end{tabular}
    
    \caption{Comparison of the spatial clustering length, $r_{0} (z)$, for  SFGs (top right) and AGNs (bottom left) with other surveys taken  from literature. The other surveys include: \citet{Overzier2003},  \citet{Dolley2014}, \citet{Magliocchetti2004},   \citealt{Gilli2005}, \citealt{Gilli2007}, \citealt{Gilli2009}, \citet{Starikova2012}, \citet{Magliocchetti2013}, \citet{Lindsay2014} \citet{Magliocchetti2017},  \citet{Hale2018}. Solid curves show the evolution of the clustering length as a function of redshift for different source types as obtained from the $S^{3}$ simulation.}
    \label{overplot_r0}
\end{figure*}

\subsection{Comparison to previous observations and discussion}

Compilation of  the  spatial clustering of AGNs and SFGs at different frequency and redshifts  from the literature and this work is summarized in Table \ref{spatial_clustering_paramateres}.
The Fig. \ref{overplot_r0}  shows  variation of the spatial clustering length   as a function of redshift. The  left and the right panels show the spatial clustering length for SFGs and AGNs respectively. Note that we only mention the spatial clustering and bias parameters of radio selected samples from previous observations in the Table \ref{spatial_clustering_paramateres}, whereas in Fig. \ref{overplot_r0} we show previous findings at radio as well as at other frequency surveys. The continuous curves gives the results from the $S^{3}$ simulation for different sub-classes of sources in the $S^3$ simulation. The individual markers with error bars corresponds to estimates of the correlation length as measured in this work and from literature.

The clustering length for SFGs found here are similar to the previous findings by  \citet{Hale2018}, \citet{Magliocchetti2017} at redshift $\textit{z} < 1$. Also, the estimated value is close to the clustering length of SFGs selected by different criteria in literature \citep{Gilli2007,Starikova2012,Magliocchetti2013,Dolley2014,Hale2018}. As mentioned in \citet{Hale2018}, the  clustering length  estimates for SFGs in radio surveys can be little higher than the other wavelength surveys, due to the fact that redshift distribution for SFGs in other wavelength surveys are skewed to lower redshifts. We have seen in Fig. \ref{AGN_SFG_distribution} that redshift distribution of SFGs shows a peak at lower redshift  and at this redshift the estimated clustering length is in good agreement with other surveys probing the same redshift range \citep{Magliocchetti2013,Hale2018}. \citet{Dolley2014} have extensively studied the redshift evolution of clustering of SFGs using mid-IR data. They have reported a slow evolution of clustering length as a function of redshift, which is also in agreement with our results. 

\begin{figure}
    \centering
     \includegraphics[width=\columnwidth,height=3in]{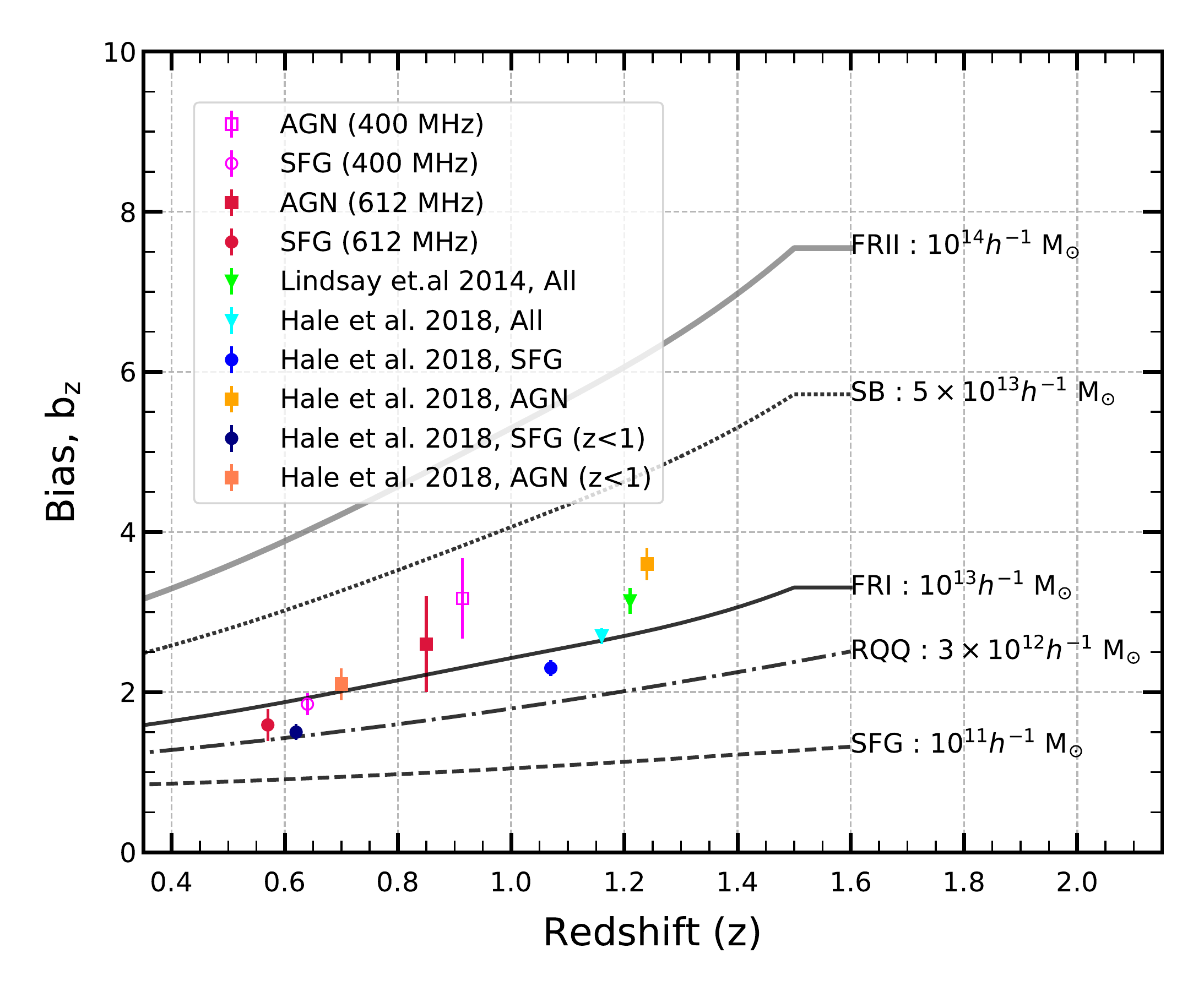}
    \caption{The bias  parameter of all sources as well as for SFGs and AGNs are shown for EN1 field as well as in other fields. We also show the bias prescription used for different source populations in SKADS simulation.}
    \label{overplot_bz}
\end{figure}

The spatial clustering length as well as bias parameter of SFGs are higher than $S^{3}$ simulation.  This may be arising for two reasons. Firstly,  
the comparison with the results from $S^3$ suggests that, if the SFGs population we study here is contaminated by a few SB galaxies, both the correlation length and the bias parameter is expected to be higher. It may also be the fact that the halo mass in the $S^3$ simulation is not a correct representation of halo mass of SFGs and is systematically lower as shown in Fig. \ref{overplot_bz}. The observational evidence of higher halo mass of SFGs in previous studies \citep{Dolley2014,Magliocchetti2017,Hale2018} suggest that the later is the likely cause of the deviation of the two parameters from the $S^3$ predictions. In such case, a comparison of  bias parameter of SFGs with bias evolution of $S^{3}$  suggests that the halo masses of SFGs at the redshift range probed here are close to $ M_{\mathrm{h}} \sim 10^{13} h^{-1} M_{\odot}$ (Fig. \ref{overplot_bz}). 

Our result on the clustering of the AGNs statistically similar to  (see Figure~\ref{overplot_r0})  the earlier reports of the clustering length estimated using radio selected \citep{Magliocchetti2017} as well as x-ray selected  \citep{Gilli2005,Gilli2009} catalogues. Comparing with the $S^{3}$ simulation suggests that the sources in our catalogue are mostly of type FRI, which agrees with the earlier results as well. Following the results of the $S^{3}$ we conclude that the AGNs are mostly hosted by massive halos of the order of  $ M_{\mathrm{h}} \sim 3-4 \times 10^{13} h^{-1} M_{\odot}$ (see Fig. \ref{overplot_bz}). This range of halo masses of AGNs are larger than that of SFGs and emphasizes that fact that AGNs inhabit more massive haloes than SFGs, hence more strong biased tracer of dark matter density field.

\section{Conclusion}
\label{sec_conclusion}
In this work,  we  investigate the  clustering properties of radio sources present in EN1 field at 400 MHz and  at 612 MHz and compare it to the the $S^3$ simulation as well as previous findings at radio and other wavelengths.  We use a luminosity based identifier for radio source populations as AGNs and SFGs. We find the angular correlation functions for all sources in our sample as well as for the SFGs and AGNs individually. The angular correlation function assumes a power law  for angular separation ranging $\sim$ $18\arcsec$ $-$  $1.5^{\circ}$. The power law index of the correlation function was found to be consistent with $1.8$, a value expected from analytical calculations \citep{Peebles1980}. Amplitude of the correlation function indicates  that the AGNs are more tightly clustered and are better tracers of the dark matter halos. 

We use the  BOSS spectroscopic survey and SWIRE photometric survey to add redshift information to the  radio sources in our catalogues and estimate the redshift distribution of the different source types.  We estimate the spatial clustering length for the sources and the corresponding bias parameters at the representative redshifts.

The clustering length and bias parameter for both samples are in excellent agreement with those found in literature from earlier observations along different lines of sight.  Comparing to the results from the $S^{3}$ simulation \citep{Wilman2008}, we found that our AGN population is mostly of type FRI and are hosted by dark matter haloes of  halo mass $  \sim 3-4 \times 10^{13} h^{-1} M_{\odot}$. We find that the halo mass used for the SFGs in the $S^3$ simulation is rather lower and a halo mass of $ M_{\mathrm{h}} \sim 10^{13} h^{-1} M_{\odot}$  is likely to host the SFGs.

It is important to note that the radio sources in our catalogue are classified based on  radio luminosity only, however, the clustering properties of  AGNs/SFGs selected based on X-ray and mid-IR data in other surveys are consistent  with our results.  We shall present results based on more robust multi-frequency identification of these sources in future.

We compare estimated parameters from our work with observations from earlier surveys. Given the statistical homogeneity and isotropy at the cosmological scales, parameter estimated  from different directions in the sky can be compared.  It is important to note that the work presented here is the first detailed investigation of the clustering properties of different types of compact radio sources in a given field in multiple frequency bands. We find the evolution of the bias and clustering length over redshift for both AGNs and SFGs as found in the EN1 field agree with earlier results from other different fields observed in radio as well as other wavelengths.

A major science objective of the  radio telescopes like LOFAR \citep{van 
Haarlem2013} or SKA \citep{Koopmans2015} is to detect the redshifted 21-cm signal from the neutral hydrogen over a large range of redshifts.  Compact extragalactic radio sources are a major source of foreground for the detection of the cosmological 21-cm signal. A major effort has been  to understand and characterise the distribution of the compact sources in the radio sky. Moreover, the 21-cm signal being very faint, the radio foreground structures need to be known to low flux density. In this work we use sub mJy source populations sources to estimate their clustering properties. This is the deepest catalogue used to estimate the foreground characteristics at frequencies where the redshifted 21 cm signal from the post-EoR epoch  is expected. Comparing our results with the prediction from $S^3$ simulation suggests that their assumption of constant halo mass bias for populating different source types is rather simplistic and needs to be reviewed. Furthermore, this emphasizes the need for further deep studies of foreground characteristics in the low frequencies. The later also would shed more light on the evolution of the clustering length and bias over redshifts and constrain structure formation models stringently.\\

{\bf ACKNOWLEDGEMENTS}

 We would like to thank the anonymous reviewer for suggestions and comments that have helped to improve this  paper. AC would like to acknowledge DST for providing INSPIRE fellowship. AC would like to thank Catherine Hale for providing the SKADS bias prescription and for many helpful suggestions. AC would like to thank Matt Jarvis for providing SKADS catalogue over private communication and also for helpful suggestions. AC would like to thank Akriti Sinha for pointing us towards BOSS catalogue for the first time.  NR acknowledges support from the Max Planck Society through the Max Planck India Partner Group grant. AD would like to acknowledge the support of EMR-II under CSIR No. 03(1461)/19.








\appendix


\bsp	
\label{lastpage}
\end{document}